\documentclass[preprint,authoryear,12pt]{elsarticle}

\usepackage{amssymb,subfig,booktabs,multirow,color,url}
\usepackage[table]{xcolor}
\usepackage{rotating,amsmath}

\journal{Computer Speech and Language}

\begin{document}

\begin{frontmatter}


\title{Data-driven Detection and Analysis \\ of the Patterns of Creaky Voice}


\cortext[cor1]{Corresponding author. Tel. +3265374749.}
\author[monsLab]{Thomas Drugman\corref{cor1}}
\ead{thomas.drugman@umons.ac.be}
\author[trinityLab]{John Kane}
\ead{kanejo@tcd.ie}
\author[trinityLab]{Christer Gobl}
\ead{cegobl@tcd.ie}

\address[monsLab]{TCTS Lab, University of Mons 31, 7000 Mons, Belgium}
\address[trinityLab]{Phonetics and Speech Laboratory,\\ School of Linguistic, Speech and Communication Sciences, \\ Trinity College Dublin, Ireland}

\begin{abstract}
This paper investigates the temporal excitation patterns of creaky voice. Creaky voice is a voice quality frequently used as a phrase-boundary marker, but also as a means of portraying attitude, affective states and even social status. Consequently, the automatic detection and modelling of creaky voice may have implications for  speech technology applications. The acoustic characteristics of creaky voice are, however, rather distinct from modal phonation. Further, several acoustic patterns can bring about the perception of creaky voice, thereby complicating the strategies used for its automatic detection, analysis and modelling. The present study is carried out using a variety of languages, speakers, and on both read and conversational data and involves a mutual information-based assessment of the various acoustic features proposed in the literature for detecting creaky voice. These features are then exploited in classification experiments where we achieve an appreciable improvement in detection accuracy compared to the state of the art. Both experiments clearly highlight the presence of several creaky patterns. A subsequent qualitative and quantitative analysis of the identified patterns is provided, which reveals a considerable speaker-dependent variability in the usage of these creaky patterns. We also investigate how creaky voice detection systems perform across creaky patterns.
\end{abstract}


\end{frontmatter}


\section{Introduction}
\label{sec:intro}

This paper presents an empirical investigation of the temporal excitation patterns related to the voice quality often referred to as creaky voice. Creaky voice is a raspy or croaking quality of the voice generally produced with a very low pitch and often with highly irregular periodicity \citep{laver80}. Creaky voice is used for a variety of functions in spoken communication and, hence, presents both an opportunity and a challenge (because of its distinctive acoustic characteristics) for speech technology.

\subsection{Terminology}
\label{sec:termin}
One major difficulty with studying creaky voice, and indeed voice quality in general, is the problem of reconciling the variation in the terminology used in the literature. Many studies use the terms \textit{irregular phonation} \citep{slifka06,surana06,vishnu06,bohm10} or \textit{glottalisation} \citep{dilley96,redi01}. However, both these terms are rather broad and indeed cover other classes of phonation (or laryngeal activity) aside from what we consider as \textit{creaky voice}. In this paper, we interpret creaky voice based solely on the auditory criterion ``a rough quality with the additional sensation of repeating impulses'' as is done in \cite{ishi08} (which is close to that in \citealp{laver80}). 
Note that the ``... sensation of repeating impulses'' part clearly discriminates the voice quality from harsh voice.
Any speech region displaying such a quality will be treated in this paper as creaky voice. Note that such a criterion will contain both creak and creaky voice, and, hence, will not apply the discrimination of the two used in \cite{laver80}. The term vocal fry (perhaps used more by American researchers; \citealp{laver80}) is often used in the literature \citep{hollien68,ishi08,wolk11}, and is likely to corresponding closely to our working definition of creaky voice.

\subsection{Creaky voice in speech communication}
\label{sec:speechComm}
Creaky voice has been studied in relation to various functions in speech communication, and most commonly with phrase or sentence boundary marking \citep{surana06_2,drugman13}. Similarly creaky voice has been associated with turn-yielding in Finnish \citep{ogden01}. However, creaky voice is likely to be also implicated in a range of speech functions other than boundary marking. It has been studied in relation to hesitations in Swedish \citep{carlson06}, and also creaky voice-like properties have been observed as an allophonic variant of word medial oral stops \citep{zue79,crystal88}. Creaky voice has also been investigated in terms of emotion and affectively coloured speech \citep{yanus05,gobl03_3,ishi08_spec} and is likely to be a significant correlate of subtle variation in levels of activation and formality of the speaking setting \citep{kanePapay11}. The use of creaky voice has also recently been shown to be increasingly common for young American females \citep{wolk11} and has also been linked to the 
portrayal 
of social status \citep{yuasa10}.

\subsection{Creaky voice in speech technology}
\label{sec:tech}
As a consequence of its implication in these various roles, the proper handling of creaky voice, and the distinctive acoustic characteristics associated with it, has a significant importance for speech technology. For speech synthesis (see e.g., \citealp{silen09,drugkane12_2}), this could result in improved naturalness for speakers who use creaky voice and also for the development of expressive speech synthesis. As it has been shown that listeners are sensitive to creaky voice in terms of recognising a speaker's identity \citep{bohm07}, it is likely beneficial for speaker recognition systems \citep{epsy06,elliot02} to exploit information to do with creaky voice. Detection of creaky voice may also benefit developments in emotion recognition and conversational analysis.

\subsection{Physiology}
\label{sec:phys}
Although the focus of this paper is on the acoustic characteristics of creaky voice, we include here a brief outline of some of the physiological characteristics reported as being associated with creaky voice. \cite{laver80} provides one of the more comprehensive descriptions of creaky voice. Here he describes creaky voice as involving low subglottal pressure, high levels of adductive laryngeal tension (i.e. the muscular tension involved in bringing the vocal folds together) and typically involves low levels of longitudinal vocal fold tension (probably the main physiological parameter utilised for pitch variation). \cite{edmond06} provide some additional physiological evidence, in particular details on the presence of {\it ventricular incursion}. This involves the ventricular folds pushing down and covering the {\it true} vocal folds, causing an increased mass. This has the consequence of lowering the frequency of vibration and often causing secondary vibrations above the glottis \citep{moisik11}.

\subsection{Acoustic characteristics of creaky voice}
\label{sec:acoustic}
The above described physiological settings have the effect of generating speech with rather distinct acoustic characteristics from modal voice\footnote{We interpret modal voice, following \cite{laver80}, as the case of periodic vocal fold vibration, with full glottal closure and no audible frication.}. However, rather than displaying a single type of acoustic pattern, speech pertaining to the auditory criterion used here can involve more than one pattern. \cite{redi01}, for instance, (following \citealp{huber88}) separate four categories of glottalisation, two of which likely correspond the auditory criterion used in the present study. The first category involves a high degree of pulse-to-pulse irregularity, in both duration (jitter) and amplitude (shimmer). This is likely to be consistent with the ``multi-pulsed'' pattern reported in \cite{ishi08,ishi07}. Note, however, that a high degree of irregularity in the duration and amplitudes of successive glottal pulses alone will not be sufficient to allow 
the ``sensation of repeating impulses''. A certain proportion of the pulses will need to display durations corresponding to a very low $F0$, 
otherwise the speech will be perceived as harsh voice \citep{laver80}.

The second pattern is categorised as having sustained low frequency with little or no superposition of formant oscillations between adjacent glottal pulses.  Such a pattern has commonly been reported, with $F0$ values typically ranging from 40-70 Hz, but at times with glottal periods as long as 100 ms \citep{blomgren98,hollien68}. Previous experiments carried out by \cite{titze94} found that human listeners begin to perceive individual pulses from around 70 Hz. This suggests that a very low $F0$, below some auditory threshold in the vicinity of 70 Hz, is sufficient to give the auditory criterion used in this paper, even in the absence of irregularity in duration or amplitude. What is unclear is the extent to which the two categories used in \cite{redi01} are overlapping. For instance, can the auditory criterion used here be achieved with highly irregular periodicity, even if many of the glottal pulses display a duration corresponding to significantly above 70 Hz?

\cite{ishi07} provide a further subdivision of the second pattern discussed by \cite{redi01}, by identifying a single-pulse and a ``double-beated'' creaky pattern, both in combination with pressed phonation. Interestingly, for the ``double-beated'' pattern, \cite{ishi07} claim that the secondary pulses observed in the speech waveform are likely caused by an abrupt glottal opening, as evidenced by the electroglottographic (EGG) signal. For this pattern, they report a very short glottal open phase and a long glottal closed phase, consistent with a tense or pressed phonation type. The same authors \citep{ishi10} report similar patterns in the phonation type known as ``Rikimi'' in Japanese. However, in that study creak is also observed within an overall laxer mode of phonation (not consistent  with the definition of ``Rikimi'') without the presence of excitations resulting from abrupt glottal openings.

Besides the temporal patterning of creaky voice, other work has reported the longest glottal closed phase for creaky voice, compared to a range of other voice qualities \citep{goblchas92}. Furthermore, the same authors reported a negative correlation between the strength of the main glottal excitation with duration of the glottal return phase, for patterns showing a high level of diplophonia.

Previous work by the present authors \citep{drugkane12_2}, sought to model the characteristics of the creaky excitation in order to achieve natural rendering of creaky voice for speech synthesis. The study reported the strong presence of secondary residual peaks corresponding to an abrupt glottal opening. The two speakers analysed in that study mainly displayed creaky voice patterns corresponding to the second category in \cite{redi01}, or the ``double-beated'' pattern in \cite{ishi07}. The data-driven excitation model derived in \cite{drugkane12_2} may therefore not be suited for the system of creaky voice involving a more irregular structure, such as for the first category patterns.

\subsection{Research aims and outline}
\label{sec:aims}
There are three main aims of the current study:

\begin{enumerate}
 \item To assess, via mutual information-based measures, the relevance of acoustic features related to creaky voice (proposed by the present authors, \citealp{drugkane12,kane13creakCSL}, as well as by others, \citealp{ishi08}) for identifying creaky voice regions.
 \item To integrate these combined features within an efficient creaky voice detection system.
 \item To utilise the classification system with different feature groups to help identify the various temporal excitation patterns of creaky voice, and to analyse them both qualitatively and quantitatively.
\end{enumerate}

The paper is structured as follows: next the speech data are described (Section \ref{sec:databases}), followed by the acoustic features included in the analysis (Section \ref{sec:features}), the mutual information of the various features is assessed in Section \ref{sec:relevance}, with the results from the automatic detection experiment in Section \ref{sec:detection}, before an examination of the creaky patterns found in the data (Section \ref{sec:patterns}) and providing a summary and a conclusion (Section \ref{sec:concl}).

\section{Databases}
\label{sec:databases}
This section describes the various speech databases used in the present study. The speech data has been selected in order to cover a variety of factors including: read vs conversational speech, different recording conditions and a range of languages.

\subsection{Text-to-speech database}
\label{sec:TTS}
The first database consists of studio recorded read speech for the purpose of text-to-speech (TTS) synthesis development. 100 sentences, which were deemed to contain creaky voice, were selected from 3 corpora. The first was speech from an American male (BDL) taken from the ARCTIC database, the second was a Finnish Female (HS) and the third was a Finnish male (MV).

\subsection{Swedish database}
\label{sec:swedish}
Recordings of a male (Swe-M) and a female (Swe-F) speaker were selected from the SPONTAL corpus of Swedish conversational speech \citep{spontal}. Each conversation lasted approximately 30 minutes and audio was captured in a recording studio. Recordings were made of audio, video and motion capture, however the data in the current study is limited to just the audio recorded with a head-mounted Beyerdynamic Opus 54 cardioid which was used to obtain optimal recording quality. 

\subsection{American database}
\label{sec:US}
The conversations were recorded involving 2 American males (US-M1 and US-M2) and 2 American females (US-F1 and US-F2)  engaged in natural dyadic conversations on the topic of food. The conversations, recorded in a quiet room, lasted around 10 minutes and audio was recorded with headset microphones.  Similar recordings were used in a recent sociological study on creaky voice \citep{yuasa10}.

\subsection{Japanese database}
\label{sec:japanese}
The final database of audio recordings consisted of conversation speech data of two Japanese female speakers (Jap-F1 and Jap-F2). The two engaged in a 30 minute conversation where they first watched some short animated films before talking carrying out the conversation. Audio was recorded on AKG C420 III PP MicroMic headset microphones wired through a BeachTek DXA-2S  pre-amp connected to the video camera (Sony DCR-TRV38 Mini DV camera).

\subsection{Creaky voice annotation}
\label{sec:annot}
It is not generally possible to obtain automatic annotation of creaky voice to the level of precision required to evaluate detection algorithms. Consequently, human annotation of creaky voice regions was carried out, closely following the procedure adopted in \cite{ishi08}. The binary decision on the presence of creaky voice was made based solely on the auditory criterion mentioned previously, i.e. ``a rough quality with the additional sensation of repeating impulses''. Annotation was, however, guided by displays of the speech waveform, spectrogram and $F0$ contours. Due to the large volume of data included in the present study, the annotation was split and shared between the first two authors who both adhered to the annotation scheme described here. Note that this annotation was achieved at a very low level, in the sense that regions of creaky regions were labelled at the frame level. We therefore expect our detection algorithms to be accurate at that level.

\subsection{Summary of speech data}
\label{sec:SummData}


Table \ref{tab:data_sum} provides a summary of the speech data used in the present study. Included in the table is a column displaying the percentage of speaking time which was annotated as involving creaky voice. It can be observed that there is a strong cross-speaker variability in terms of the proportion of speech containing creaky voice. One American speaker used creaky voice in 3.6 \% of her speech, while another American female, recorded as part of the same database, used creaky voice in over 10 \% of her speech. Also included in Table \ref{tab:data_sum} is the actual duration (in seconds) of creaky productions for each speaker. Note in the TTS data the Finnish sentences are considerably longer than the US English sentences. Also, in the conversational data the duration given is the time they participated in the conversation, where as the percentage of creak is determined from just those times where the person is actually speaking. For instance, speaker Jap-F2 speaks less than Jap-F1 over the course of 
a 30 minute conversation, but a larger proportion of her speaking time involved creaky phonation.

\begin{table} 
\caption{{Summary of the speech data used in this study.}}
\vspace{-.4cm}
  \begin{center}
\vspace{0cm}
\label{tab:data_sum}
    \begin{tabular}{llcccccc}
\toprule
{\bf Database} & {\bf ID} & \bf{Gender} & \bf{Country}  & \bf{Duration} & \bf{Creak (\%)} & \bf{Creak duration (s)}  \\
\midrule
\multirow{3}{*}{{\bf TTS}} & BDL & Male & USA &  100 Sentences & 7.6 & 19.39\\
& HS & Female & Finland &  100 Sentences & 5.8 & 36.86\\
& MV & Male & Finland &  100 Sentences & 7.1 & 31.56\\
\midrule
\multirow{2}{*}{{\bf Swedish}} & Swe-F & Female & Sweden & 30+ Minutes & 5.3 & 64.13 \\
& Swe-M & Male & Sweden &  30+ Minutes & 5.8 & 32.13 \\
\midrule
\multirow{4}{*}{{\bf US}} & US-F1 & Female & USA &  10+ Minutes & 10.5 & 32.70\\
& US-M1 & Female & USA &  10+ Minutes & 7.4 & 15.87 \\
& US-M2 & Male & USA &  10+ Minutes & 9.2 & 16.00 \\
& US-F2 & Male & USA & 10+ Minutes & 3.6 & 9.09 \\
\midrule
\multirow{2}{*}{{\bf Japan}} & Jap-F1 & Female & Japan &  30+ Minutes & 4.8 & 55.68\\
& Jap-F2 & Female & Japan & 30+ Minutes & 6.5 & 40.71\\
\bottomrule
    \end{tabular}
  \end{center}
\vspace{0cm}
\end{table}

\section{Features for Creaky Voice Characterisation}
\label{sec:features}
This section describes the set of acoustic features relevant to creaky voice which are examined in the present study. This set consists of features developed by the present authors \citep{kane13creakCSL,drugkane12} as well as features by \cite{ishi08}. A brief description of each of the features is given below.

\subsection{H2-H1 and F0creak}
\label{sec:H2H1}
\begin{figure}[ht!]
  \begin{center}
    \includegraphics[width=15cm]{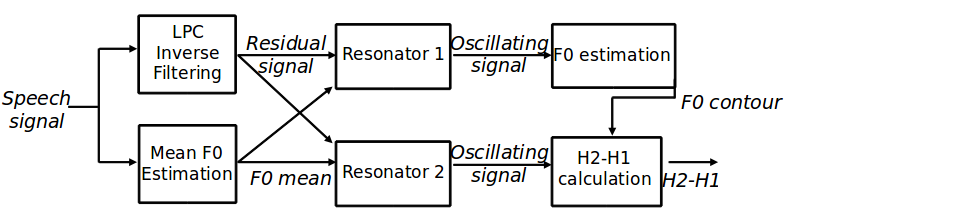}
  \end{center}
\caption{{{ Block diagram of the H2-H1 parameter, derived from the response of two resonators excited by the LP-residual signal. Both resonators are centred on $f_{0,mean}$ but use different bandwidths. For details see text.}} 
\label{fig:H2H1_workflow}}
\end{figure}

The first acoustic feature was originally presented in \cite{drugkane12}, and is designed to characterise the strong presence of secondary residual peaks often found in creaky voice. The block diagram (in Figure \ref{fig:H2H1_workflow}) shows that two resonators are applied to the linear prediction (LP) residual. Both resonators have a centre frequency set to the speaker's mean $F0$, however each has different bandwidth settings. Resonator 1 is set with a bandwidth of around 1 kHz and is used for providing a more robust $F0$ contour, even in creaky voice regions. Note that the $F0$ contour is derived by calculating a corrected autocorrelation function, $r'(\tau)$, from 50-ms Hanning-windowed frames of the Resonator 1 output:

\begin{equation}
 \label{eq:autocorr}
r'(\tau) = \frac{N}{N-\tau}\cdot \mbox{autoCorr}(\tau)
\end{equation}

\noindent where $N$ is the window length (in samples) and $\tau$ is the number of autocorrelation lags. A correction of $\frac{N}{N-\tau}$ is applied to compensate for the decreasing properties of the autocorrelation function with increasing $\tau$ (as is used in \citealp{ishi08}). The local fundamental period is then considered as the position of the maximum value in $r'(\tau)$ above the peak centred on $\tau = 0$.

Resonator 2 is set with a bandwidth to 150 Hz, and is used for calculating the H2-H1 feature. Again applying 50-ms frames Hanning-windowed frames, this time to Resonator 2 output, the amplitude spectrum is derived and the corresponding $F0$ value (derived from Resonator 1) is used to detect the amplitudes of the first two harmonics (H1 and H2). The difference in amplitude of these two harmonics (i.e. H2-H1) in dB is the first acoustic feature used in this study. The H2-H1 contour is smoothed using a 100-ms moving average filter to remove the effect of outliers.

\begin{figure}[ht!]
  \begin{center}
   \includegraphics[width=13cm]{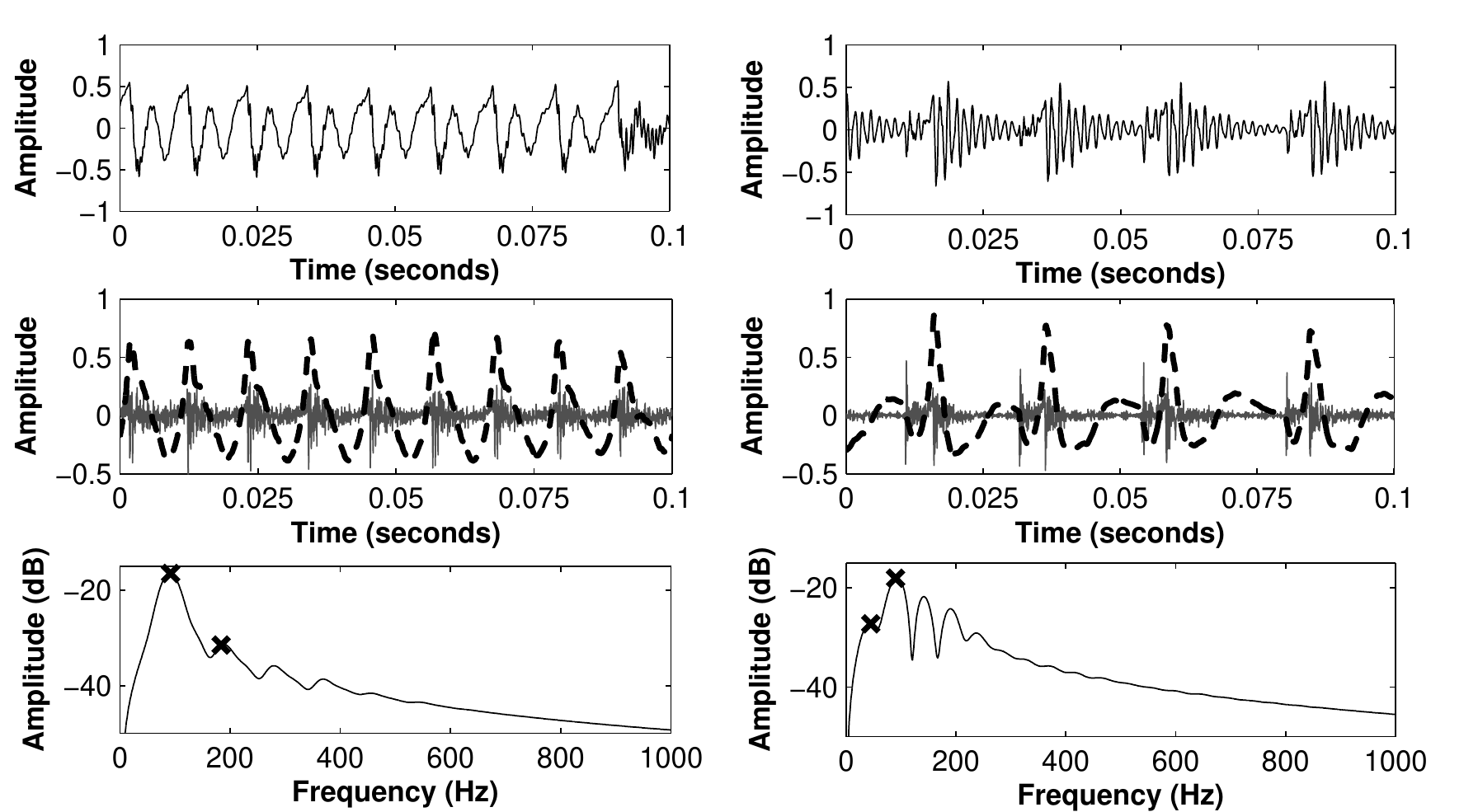}
  \end{center}
\vspace{0cm}
\caption{{ Example of modal phonation (left column) and creak (right column) uttered by the same speaker, with the speech waveform (top row),  the LP residual signal (middle row, solid line) together with the output of Resonator 2 (in dashed line), and  the amplitude spectrum (in dB) of a frame of the output of Resonator 2 where the values for $F0$ and $2\cdot F0$ are indicated by crosses (bottom row).}} 
\label{fig:H2H1_illustration}
\vspace{0cm}
\end{figure}

An illustration of the steps involved in the H2-H1 calculation process is given in Figure \ref{fig:H2H1_illustration}. On left side, for modal phonation, one can observe no strong secondary peaks in the residual signal or in the Resonator 2 output (middle panel). Consequently, its amplitude spectrum (bottom panel) shows a prominent amplitude level at $F0$ compared to the second harmonic. Contrastingly, for a creaky voice segment (right side), strong secondary peaks can be observed in the residual signal (middle panel) which affects the Resonator 2 output, causing a greater harmonicity in the resulting 
amplitude spectrum (bottom panel). A combination of this increased harmonicity and a considerably lower $F0$ in creaky voice compared to the speaker's mean $F0$, imply that the amplitude of the first harmonic is considerably weaker to that of the second. Note that in this study both the H2-H1 and the $F0$ contour derived from Resonator 1 (named F0creak) are used in this study.

One limitation of the H2-H1 parameter is that for highly irregular periodicity patterns (i.e. the category 1 \citep{redi01}, multi-pulsed pattern \citep{ishi08}), the amplitude spectrum may not display any clear harmonics in which case the discriminating power of H2-H1 may be reduced.

\subsection{Residual peak prominence - Peak-Prom}
\label{sec:res_p}
The second feature (Peak-Prom) was designed to avoid spectral and periodicity related measurements, and instead characterise each excitation peak in the time-domain. The Peak-Prom parameter essentially characterises the prominence of LP-residual peaks relative to its immediate neighbourhood. The output of Resonator 1 (see Figure \ref{fig:H2H1_workflow}) is used, as its low-pass filtering effect makes the prominence measure more robust as opposed to measurements directly from the LP-residual. Furthermore, the large bandwidth (1 kHz) ensures a rapid decay in the oscillations which facilitates the prominence measurement. Peak-Prom involves the use of a fixed, non-overlapping rectangular window whose duration is set to 30 ms. This roughly corresponds to two periods at 70 Hz. In this method correct polarity of the speech signal is assumed (this can be determined automatically for example using the method described in \citealp{drugmanPol}). Although with correct polarity the LP-residual displays positive peaks, 
the low-pass filtering effect of the resonator causes corresponding negative peaks in its output. The resonator output is inverted so that it instead displays strong positive peaks.

For each frame the absolute maximum peak in the resonator output is identified and the frame is then shifted to be centred on this peak. By measuring the amplitude difference between the maximum peak (in the centre of the frame) and the next strongest peak one can obtain a parameter value which differentiates modal and creaky regions. In order to avoid selecting values in the vicinity of the main centre peak, the search for the next strongest peak is made outside a distance of 6 ms on both sides of the centre of the frame.
This corresponds to 40 \% of half the frame length which ensures that there is sufficient space for peaks to occur from neighbouring glottal pulses. A value is thus obtained for each frame producing the outputted parameter contour. This contour is then filtered with a 3-point median filter to remove misdetections due to transients in the signal. 

The Peak-Prom parameter calculation can be summarised in the following steps:

\begin{enumerate}
 \item Apply a 30 ms rectangular window to the Resonator 1 output
 \item Invert the windowed frame to ensure positive peaks
 \item Identify the maximum peak and shift the frame to be centred on this peak
 \item Measure the next strongest peak outside the middle 12 ms of the frame, and calculate the amplitude difference between it and the strongest peak
 \item Repeat with non-overlapping frames
 \item Apply 3-point median filter to the extracted contour
\end{enumerate}

\begin{figure}[htp]
  \begin{center}
   \includegraphics[width=12cm]{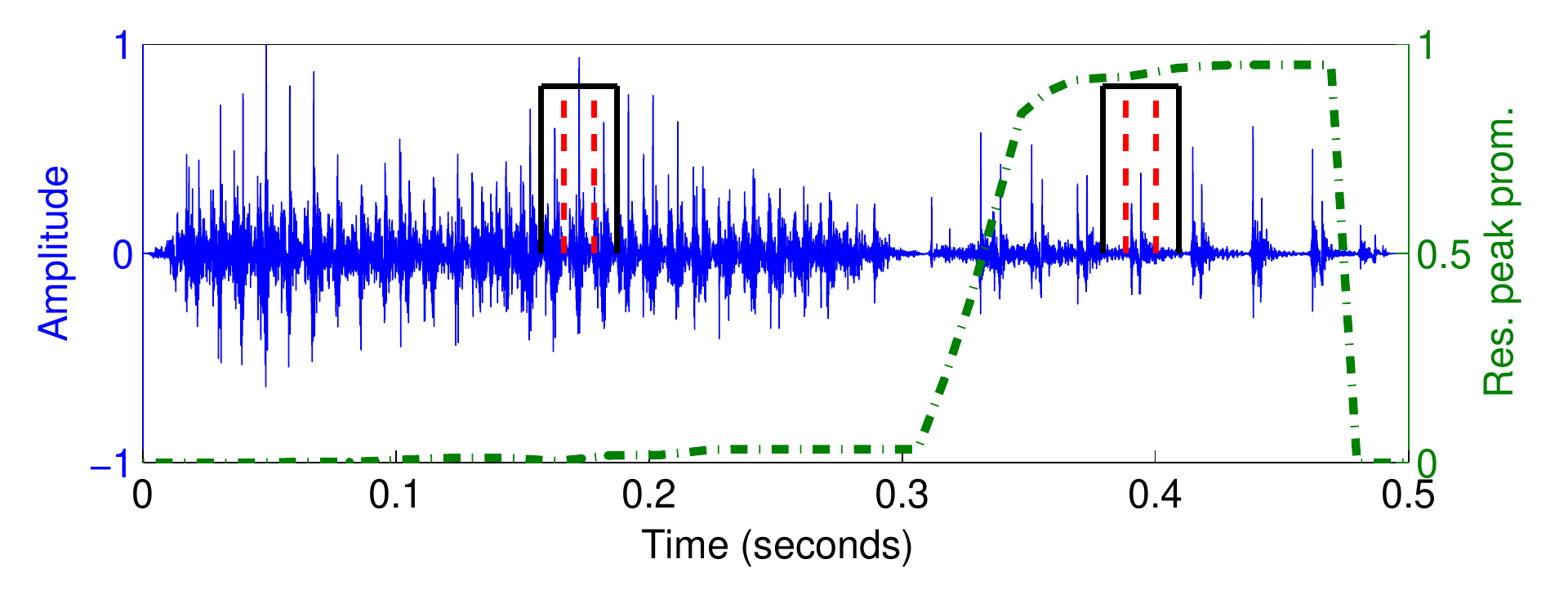}
  \end{center}
\vspace{0cm}
\caption{{ Linear Prediction residual of an utterance with creaky voice starting from around 0.3 seconds, along with the Peak-Prom contour (green dash-dotted line). Also shown are sample windowing locations (black line) and and exclusion areas (red dashed line) for the parameter calculation in both the modal and creaky region.}} 
\label{fig:res_p_example}
\vspace{0cm}
\end{figure}

An illustration of the Peak-Prom parameter is shown in Figure \ref{fig:res_p_example}. An example of the window (black line) is provided both in the modal voice region (first syllable) and in the creaky voice region (second syllable). The dashed red line shows the region within the window from which the second peak measurement is excluded. For the creaky voice window it can be noticed that aside from the prominent excitation peak in the centre of the frame, no other strong peaks can be observed (outside the exclusion region). As a result the Peak-Prom contour (dot-dashed green line) displays high values here. For the modal voice region, one can observe that several strong residual peaks are contained within the rectangular frame and, hence, Peak-Prom displays values close to 0. A limitation of the Peak-Prom feature is that if a given creaky voice pattern (corresponding to the category 1 \citep{redi01}, multi-pulsed pattern \citep{ishi08}) contains a high proportion of glottal pulses with a duration 
significantly lower than 15 ms, then 
the effectiveness of Peak-Prom may be reduced.

Note that for the present study the features H2-H1, F0creak and Peak Prom are grouped as the \textbf{KD} (Kane-Drugman) features.

\subsection{Power peak parameters - PwP fall \& PwP rise}
\label{sec:pwp}
\begin{figure}[ht!]
  \begin{center}
   \includegraphics[width=13cm]{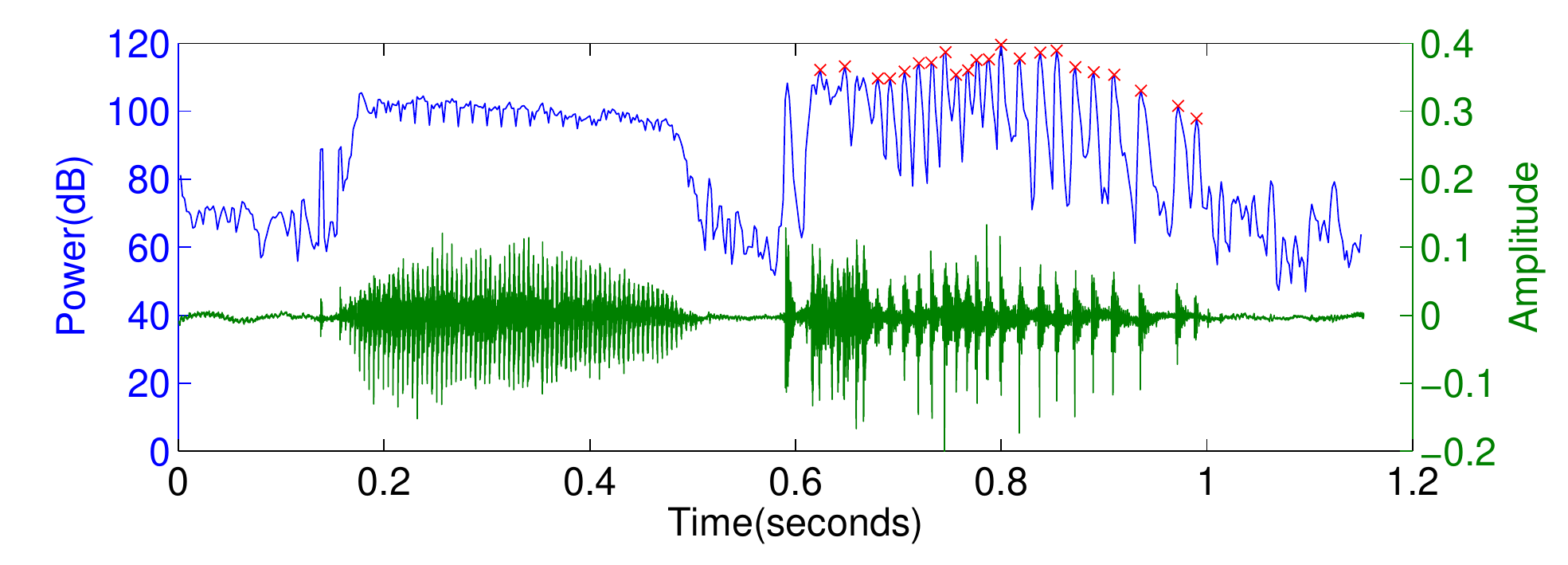}
  \end{center}
\vspace{-.5cm}
\caption{Very short-term power contour (blue contour), with power peaks (red stars) and the speech waveform (green) of the utterance ``very good'' with creaky voice on ``good'' spoken by an American female during natural conversation.} 
\label{fig:pwp}
\vspace{0cm}
\end{figure}

The first features described in \cite{ishi08} used in this study are the so-called Power-Peak (PwP) parameters. Note that all of the patterns from \cite{ishi08} are calculated from the speech signal bandlimited to 100-1500 Hz. A `very short-term' power contour is measured, with a frame length of 4 ms and shift of 2 ms, in order to highlight the amplitude variation within individual pulses (see Figure \ref{fig:pwp}). Peaks are then detected in this contour and Power Peak (PwP) parameters are derived for each peak based on the previous (PwP-rising) and following (PwP-falling) 5 frames (i.e. 10 ms) in the contour. The maximum power difference in each direction is used as the PwP value. In the original implementation a threshold is applied to this parameter to determine whether the peak can be used as a creak candidate location, however in the present study we simply use both PwP parameter values as acoustic features.

\subsection{Inter-pulse similarity - IPS}
\label{sec:IPS}
\begin{figure}[ht!]
  \begin{center}
   \includegraphics[width=13cm]{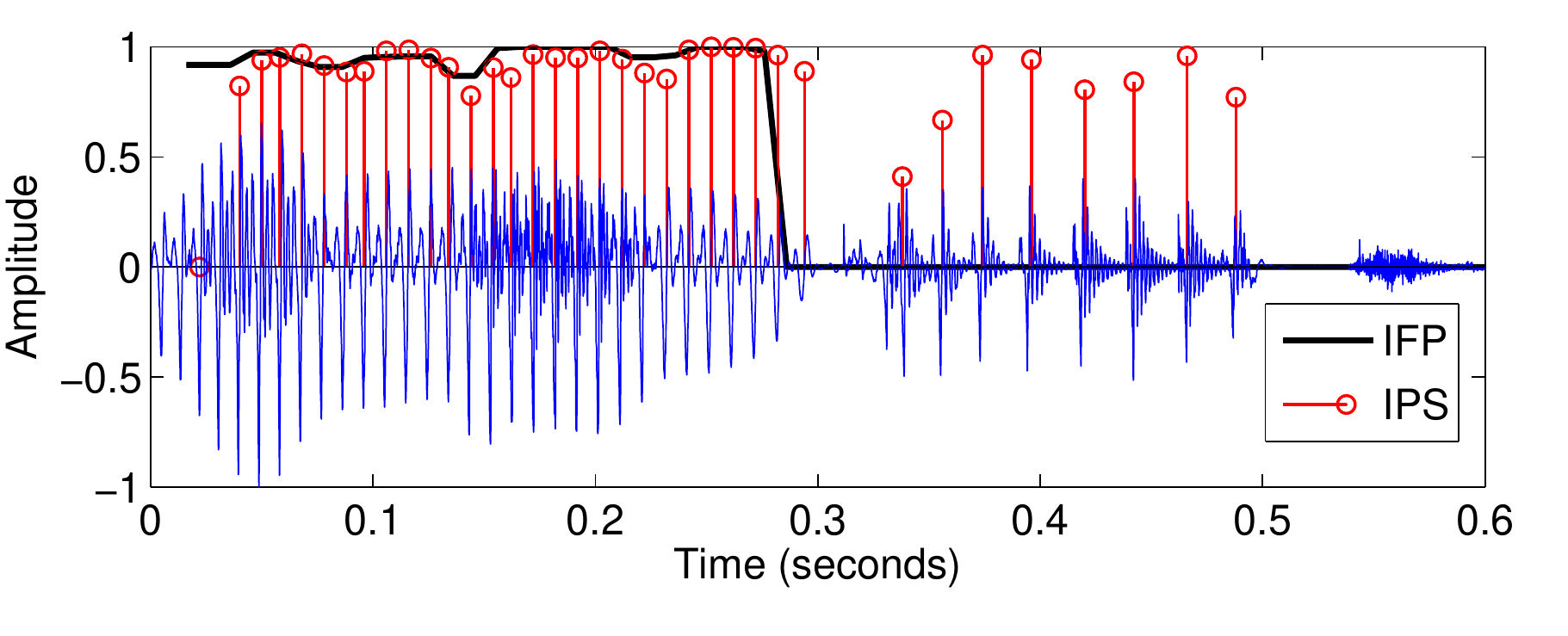}
  \end{center}
\vspace{-.5cm}
\caption{Speech waveform along with the IFP contour (black line) and the IPS values (red stems) of the utterance ``... remember it'' with creaky voice on the syllable ``it'' spoken by an American male.} 
\label{fig:IFP_IPS}
\vspace{0cm}
\end{figure}

The inter-pulse similarity measure \citep{ishi08} is used to discriminate glottal pulses corresponding to creaky voice from unvoiced regions. The parameter is derived using the locations of the peaks measured in the very short-term energy contour (Section \ref{sec:pwp}). A cross-correlation function is applied to assess the similarity of adjacent bandlimited speech pulses:

\begin{equation}
 \label{eq:ccorr}
\mbox{IPS} = \max\left\{\mbox{CCorr}(F_{\tau_1},F_{\tau_2}); \quad \tau_1-\tau_2 < T_{max}  \right\}
\end{equation}

\noindent where CCorr is the cross-correlation function, $F_{\tau_1}$ and $F_{\tau_2}$ are the frames centred on successive candidate peak locations, and $T_{max}$ is the maximum allowed distance between adjacent peaks, and is set to 100 ms. Each frame is selected as the range of 5 ms around the peak location. It is assumed that adjacent creaky voice pulses will show a high degree of similarity, as the vocal tract is unlikely to have significantly changed in such a short space of time and, hence IPS values should be high. Contrastingly, for unvoiced regions IPS values should be low, indicating a low level of similarity. The IPS parameter is illustrated in Figure \ref{fig:IFP_IPS}, where the values are shown as red stems. It can be observed that high IPS values are found in both the `modal' voice region (up to 0.3 second) and also in the creaky voice (0.33 to 0.5 seconds).

Note that both the IPS and PwP features are sampled on a glottal synchronous basis. These features are interpolated up to a fixed frame basis using a nearest neighbour approach for the current study.

\subsection{Intra-frame periodicity - IFP}
\label{sec:IFP}
The Intra-frame periodicity (IFP) feature \citep{ishi08} was originally designed to help disambiguate creaky voice from other voiced regions. Unlike the IPS feature, IFP is calculated on a fixed frame basis using:

\begin{equation}
 \label{eq:IFP}
\mbox{IFP} = \min\left\{\frac{N}{N-\tau}\cdot \mbox{autoCorr}(\tau); \quad \tau = j\cdot \tau_0; \quad j = 1, 2, ... \right\} 
\end{equation}

\noindent where $N$ is the frame length (set to 32 ms, with a 10 ms shift), $\tau$ is the autocorrelation lag, autoCorr is the normalised autocorrelation function, and $\tau_0$ is the lag of the strongest autocorrelation peak. Note also that the search space for $\tau$ is limited to 15 ms and that the factor $\frac{N}{N-\tau}$ is used to compensate for the decrease in amplitude with increasing $\tau$ in the autocorrelation function. Due to irregular periodicity and/or the very low $F0$ of creaky voice, IFP values in these regions will be close to 0 (see the last syllable in Figure \ref{fig:IFP_IPS}). Other voiced regions (for instance the first syllable in Figure \ref{fig:IFP_IPS}) will display IFP values close to 1.

Although in many cases IFP is suitable for discriminating creaky voice from non-creaky voiced regions, its effectiveness for this purpose can be reduced somewhat when a speech region contains a very low $F0$ (but not quite sufficiently low for creaky voice, e.g. around 80 Hz). Such effect was observed in a previous study \citep{drugkane12} and resulted in a high number of false detections for a speaker with an inherently low pitch.

Note that in the present study the features PwP-fall, PwP-rise, IPS and IFP are given the group title \textbf{Ishi's} features, as they were all proposed in \citep{ishi08}.

\subsection{Additional acoustic features - Energy norm, Power std \& ZeroXrate}
\label{sec:add_feat}
In addition to the acoustic features (KD) proposed by the present authors and those by \cite{ishi08}, three further features were included in particular to avoid false positives in unvoiced and silent regions. It was suggested (in a personal communication) by the authors of \cite{ishi08} that creaky voice detected in regions of considerably lower energy (e.g., 20 dB) than the maximum energy of an utterance, be discarded from the detection output. Consequently, we include a measure of signal energy (in dB) which has been normalised to the maximum energy of the utterance (Energy Norm). We also include ZeroXrate, which is a measure of the number of zero-crossings per ms. Unvoiced and silent regions are likely to display a significantly higher rate of zero-crossings compared with creaky voice regions \citep{kanePapay11}. Note that both Energy Norm and ZeroXrate are measured on 32 ms frames, with a 10 ms shift. Finally, the Power Std feature is used as a measure of the variance in the very short-term power 
contour used in the calculation of the PwP features \citep{ishi08}. The feature is derived as the standard deviation of the power (in dB), measured on 16 frames (corresponding to 32 ms). As with the other features used in this study, these three features were sampled every 10 ms.

\section{Mutual Information-based Relevance of the Creaky Features}
\label{sec:relevance}
The goal of this section is to assess the relevance of the features described in Section \ref{sec:features} for the automatic detection of creaky regions. For this, we here make use of measures derived from Information Theory \citep{Shannon} as they allow a quantification of the amount of discriminant information conveyed by the features. This is done independently of any subsequent classifier. Our approach allows for an integrated assessment of the discriminative power of each feature individually and also of the discriminative power in the context of the other features in terms of redundancy and synergy.  These measures are first presented in Section \ref{ssec:MImeasures}. They are then used for an objective assessment of the features in Section \ref{ssec:MIresults}.

\subsection{Mutual Information-based Measures}
\label{ssec:MImeasures}

The problem of automatic classification consists of finding a set of features $X_i$ such that the uncertainty on the determination of classes $C$ is reduced as much as possible \citep{FSBook}. For this, Information Theory \citep{Cover} allows the assessment of the relevance of features for a given classification problem, by making use of the following measures (where $p(.)$ denotes a probability density function):

\begin{itemize}
\item The entropy of classes $C$ is expressed as:
\begin{equation}
H(C)=-\sum_c{p(c)\log_2p(c)}
\label{eq:entropy}
\end{equation}
where $c$ are the discrete values of the random variable $C$. $H(C)$ can be interpreted as the amount of uncertainty on the class determination.

\item The mutual information (MI) between one feature $X_i$ and classes $C$ is defined as:
\begin{equation}
I(X_i;C)=\sum_{x_i}{\sum_c{p(x_i,c)\log_2\frac{p(x_i,c)}{p(x_i)p(c)}}}
\label{eq:MI}
\end{equation}
where $x_i$ are the discretised values of feature $X_i$. $I(X_i;C)$ can be viewed as the information that feature $X_i$ conveys about the considered classification problem, i.e. the intrinsic  discrimination power of a given feature.

\item The joint mutual information between two features $X_i$, $X_j$, and classes $C$ can be expressed as:
\begin{equation}
I(X_i,X_j;C)=I(X_i;C)+I(X_j;C)-I(X_i;X_j;C)
\label{eq:jointMI}
\end{equation}
and corresponds to the information that features $X_i$ and $X_j$, when \emph{used together}, bring to the classification problem. The last term can be written as \citep{Cover}:
\begin{equation}
\begin{split} I(X_i;&X_j;C)= \\
\sum_{x_i}\sum_{x_j}\sum_cp(x_i,x_j,c)\cdot&\log_2\frac{p(x_i,x_j)p(x_i,c)p(x_j,c)}{p(x_i,x_j,c)p(x_i)p(x_j)p(c)}
\end{split}
\label{eq:redundancy}
\end{equation}
\end{itemize}

An important remark has to be underlined about the sign of this term. It can be noticed from Equation \ref{eq:jointMI} that a positive value of $I(X_i;X_j;C)$ implies some \textit{redundancy} between the features, while a negative value means that features exhibit some \textit{synergy} (depending on whether their association brings respectively less or more than the addition of their own individual information).

To evaluate the significance of the features described in Section \ref{sec:features}, the following measures are computed:

\begin{itemize}
\item the \textit{relative intrinsic information} of one individual feature $\frac{I(X_i;C)}{H(C)}$, i.e. the proportion of relevant information conveyed by the feature $X_i$,
\item the \textit{relative redundancy} between two features $\frac{I(X_i;X_j;C)}{H(C)}$, i.e. the rate of their common relevant information,
\item the \textit{relative joint information} of two features $\frac{I(X_i,X_j;C)}{H(C)}$, i.e. the proportion of relevant information they convey together.
\end{itemize}

For this, Equations \ref{eq:entropy} to \ref{eq:redundancy} are calculated. Probability density functions are estimated by a histogram approach using bins uniformly distributed between the possible extremum values. The number of bins is set to 50 for each feature dimension, which results in a trade-off between an adequately high number for an accurate estimation, while keeping sufficient samples per bin. Class labels correspond to the presence ($c=1$) or not ($c=0$) of a creakiness in the voice, as indicated by the manual annotation.

\subsection{Results}
\label{ssec:MIresults}

\begin{figure}[ht!]
  \begin{center}
   \includegraphics[width=14cm]{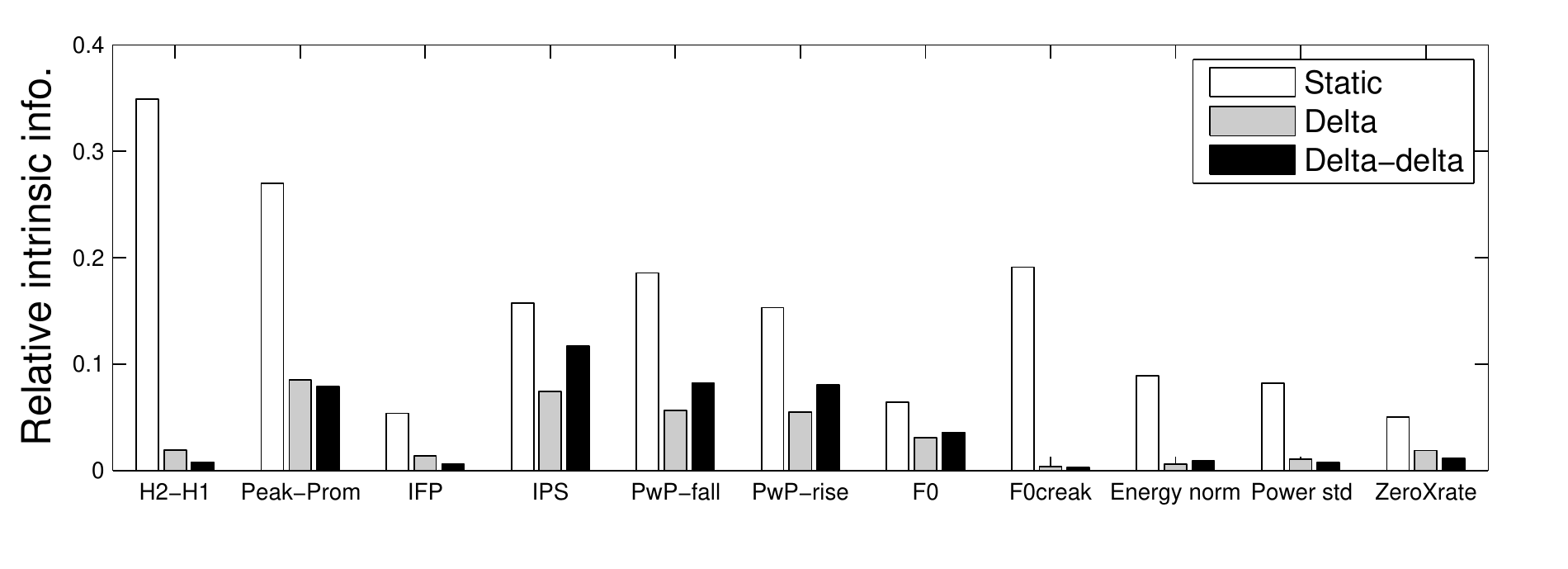}
  \end{center}
\vspace{-.6cm}
\caption{Intrinsic mutual information of the features designed for creaky voice detection. The first and second derivatives of these features are considered in addition to their static values.} 
\label{fig:MIvalues}
\vspace{0cm}
\end{figure}

First, the intrinsic discrimination power of each feature is investigated. Figure \ref{fig:MIvalues} displays the relative intrinsic information for the features described in Section \ref{sec:features} as well as for their first and second derivatives. Interestingly, the most informative features are H2-H1 (around 0.35) and Peak-Prom (about 0.27), while the best of Ishi's features is \emph{PwP fall} with 0.19. The need to develop a $F0$ tracker specific to the analysis of creaky voice is clearly emphasised, as \emph{F0creak} reaches 0.19 when \emph{F0} extracted with a standard \textit{F0} tracker only reaches 0.07. The $F0$ tracker used here is the summation of residual harmonics (SRH) algorithm \citep{drugman11}. Note that both the low fundamental frequency and irregular temporal characteristics of creaky voice are known to cause trouble for $F0$ trackers. This was highlighted in a recent study \citep{raitioCreakSynth} which demonstrated that the majority of $F0$ trackers incorrectly determined a 
significant proportion of creaky voice regions to be unvoiced.

It turns out that the dynamics of the features for creaky voice characterisation only conveys a limited amount of information compared to their static version. Finally, it is worth noting that the widely-used Mel-frequency cepstral coefficients (MFCCs) achieve, at best, a relative intrinsic information of only 0.04 and their usefulness for creaky voice detection is therefore negligible.

Although Figure \ref{fig:MIvalues} gives some answers regarding the discrimination power of each feature independently, nothing is said about their possible complementarity or redundancy. This aspect is now tackled in Table \ref{tab:MItable} for the KD and Ishi's features (which were clearly among the most informative in Figure \ref{fig:MIvalues}). The best combination of two features is H2-H1 with Peak-Prom (0.48). Nonetheless it is of interest to observe that the association of H2-H1 with Ishi's features (except IFP) carries out comparable results (between 0.46 and 0.47), albeit that these latter features, individually, display less intrinsic information. This is possible thanks to the weak redundancy of Ishi's features with H2-H1 (varying only between 0.03 and 0.06). This good complementarity can then be turned into an advantage for building up an automatic detection system. One should note that Ishi's features, as opposed to the KD features, were not developed to be individual detectors of creaky voice, 
but 
rather their combination is what is relevant in this regard (i.e. detecting candidate creaky regions, and eliminating normal voiced and unvoiced candidate regions). This is evident here by the negative value (-0.06) for relative redundancy between IFP and IPS, indicating a degree of synergy between the two parameters. However, aside from this pair features inside a category (i.e. KD or Ishi's features) are quite redundant with each 
other while features from two different categories exhibit an interesting complementarity.

\begin{table}[htbp]
     \centering    
\begin{tabular}{|c|c|ccc|cccc|}
\hline
& & \multicolumn{3}{c|}{ KD  } & \multicolumn{4}{c|}{ ISHI'S  }\\
\hline
& & \rotatebox{90}{H2-H1} & \rotatebox{90}{Peak-Prom} & \rotatebox{90}{F0creak} & \rotatebox{90}{IFP} & \rotatebox{90}{IPS} & \rotatebox{90}{PwP\_fall} & \rotatebox{90}{PwP\_rise}\\
\hline
\multirow{3}*{\rotatebox{90}{{\footnotesize {KD}}}}
& H2-H1 & \cellcolor{orange!60}\textbf{0.35} & 0.48 & 0.38 & 0.37 & 0.46 & 0.47 & 0.46\\
& Peak-Prom & \cellcolor{orange!15}0.14 & \cellcolor{orange!60}\textbf{0.27} & 0.37 & 0.31 & 0.37 & 0.37 & 0.36\\
& F0creak & \cellcolor{orange!15}0.16 & \cellcolor{orange!15}0.09 & \cellcolor{orange!60}\textbf{0.19} & 0.21 & 0.29 & 0.30 & 0.27\\
\hline
\multirow{4}*{\rotatebox{90}{{\footnotesize ISHI'S}}}
& IFP & \cellcolor{orange!15}0.03 & \cellcolor{orange!15}0.02 & \cellcolor{orange!15}0.03 & \cellcolor{orange!60}\textbf{0.05} & 0.27 & 0.24 & 0.21\\
& IPS & \cellcolor{orange!15}0.04 & \cellcolor{orange!15}0.06 & \cellcolor{orange!15}0.06 & \cellcolor{orange!15}-0.06 & \cellcolor{orange!60}\textbf{0.16} & 0.23 & 0.22\\
& PwP\_fall & \cellcolor{orange!15}0.06 & \cellcolor{orange!15}0.08 & \cellcolor{orange!15}0.08 & \cellcolor{orange!15}0.00 & \cellcolor{orange!15}0.11 & \cellcolor{orange!60}\textbf{0.19} & 0.21\\
& PwP\_rise & \cellcolor{orange!15}0.05 & \cellcolor{orange!15}0.06 & \cellcolor{orange!15}0.07 & \cellcolor{orange!15}0.00 & \cellcolor{orange!15}0.09 & \cellcolor{orange!15}0.13 & \cellcolor{orange!60}\textbf{0.15}\\
\hline
\end{tabular}
\caption{Mutual information-based measures for the best creaky features. \emph{On the diagonal - orange}: the relative intrinsic information $\frac{I(X_i;C)}{H(C)}$. \emph{In the bottom-left part - light orange}: the relative redundancy between the two considered features $\frac{I(X_i;X_j;C)}{H(C)}$. \emph{In the top-right part - white}: the relative joint information of the two considered features $\frac{I(X_i,X_j;C)}{H(C)}$.}
\label{tab:MItable}
\end{table}

\begin{figure}[ht!]
  \begin{center}
   \includegraphics[width=13cm]{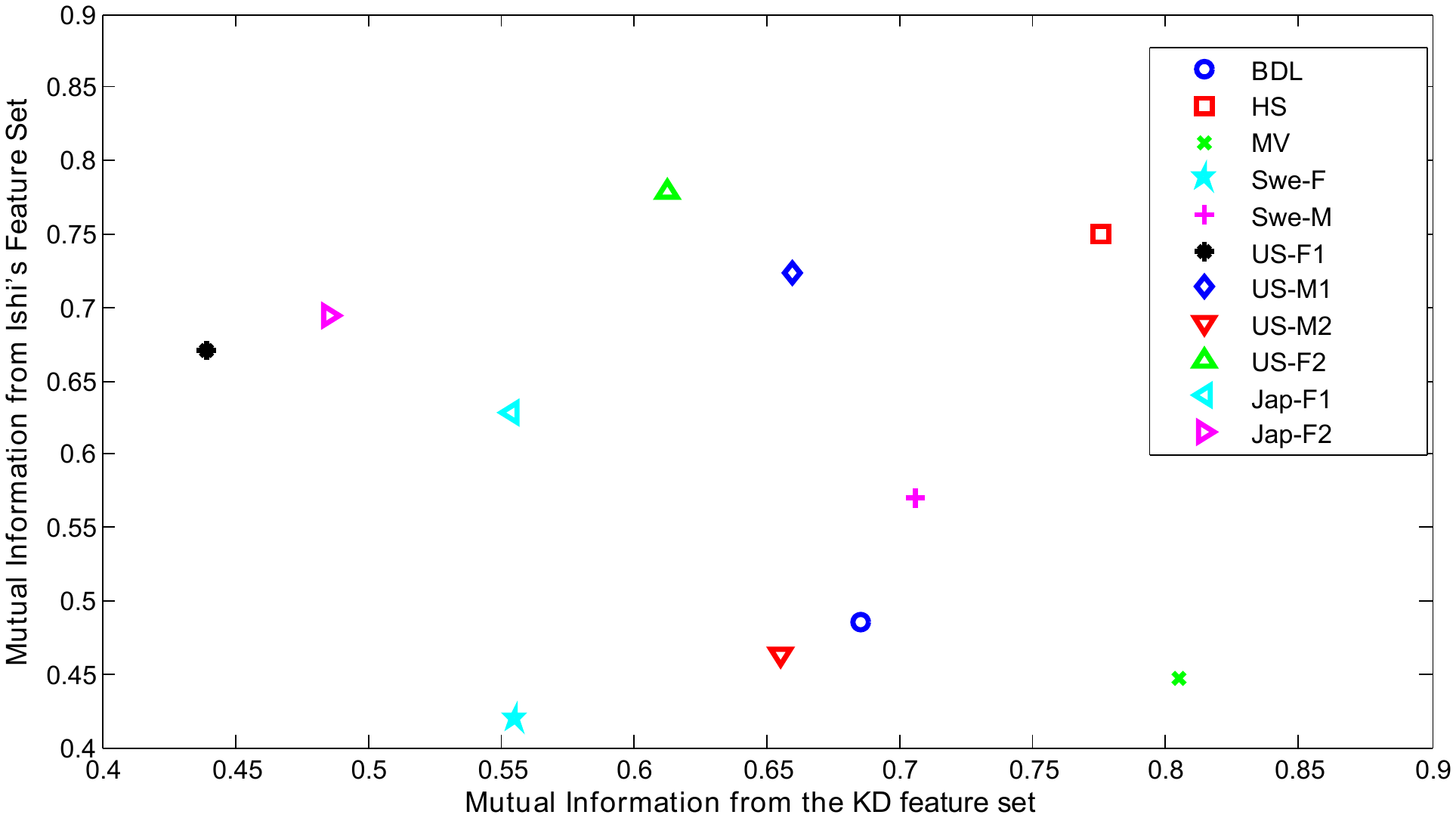}
  \end{center}
\vspace{-.6cm}
\caption{Mutual Information brought by the two sets of features for the various speakers of the database.} 
\label{fig:FinalMIProfile}
\vspace{0cm}
\end{figure}

The good complementarity between the two groups of features indicates that they possibly reflect different characteristics of the creaky production and might therefore be linked to the realisation of different creaky patterns. Figure \ref{fig:FinalMIProfile} investigates the speaker-dependent variability of the usefulness of both the KD and Ishi's features. Note that to calculate the total MI for a set of features, we cope with redundancy as suggested in \cite{DrugmanFS}. Except HS for whom both groups of features seem to provide a comparable amount of information, it seems that speakers tend to predominantly use a creaky production well characterised either by the KD or by Ishi's features. This observation tends to reinforce the possible existence of at least two creaky patterns as described by the two groups of features.

\section{Automatic Detection of Creaky Voice}
\label{sec:detection}

This section aims at the integration of the acoustic characteristics described in \ref{sec:features} within a system for the automatic detection of creaky voice. This system is first explained in Section \ref{ssec:ANN} and the experimental protocol is presented in Section \ref{ssec:ExpProtocol}. The results of classification are given in Section \ref{sec:DetectionResults} which further strengthens our findings from the MI-based analysis described in Section \ref{ssec:MIresults}.

\subsection{Proposed System for Automatic Detection of Creaky Voice}
\label{ssec:ANN}

In order to automatically detect the creaky regions from an audio stream, the features described in Section \ref{sec:features} have been used together with their first and second derivatives to account for their dynamics. More precisely, three sets of features have been considered: Ishi's features, KD features, and the combination of both approaches. In the 3 cases, characteristics informative about audio activity (\emph{Energy\_Norm}, \emph{Power\_std} and \emph{ZeroXrate}) have been appended to the feature vector in order to avoid spurious detections during silences or in sudden bursts (which may frequently occur in conversational data).

Relying on these features, two types of classifier were trained: a Binary Decision Tree (BDT) and an Artificial Neural Network (ANN). In the BDT, the separation of the two classes is done using a top-down approach where both classes are initially placed at the root node and then a series of binary questions are asked (to do with the input features) and for each question a new child node is created. This builds up the decision tree, the ends of which are leaf nodes. The commonly used Gini's Diversity Index (GDI) was employed for the splitting criterion and splits are selected in order to reduce the GDI criterion. The splitting was stopped if the current node was \textit{pure} (i.e. contained only observations of a single class) or if the branch node contains fewer than 10 observations. Note that the KD features used with the BDT classifier was proposed in our previous work \citep{kane13creakCSL}, where it was shown to outperform the original Ishi's creaky voice detection system \citep{ishi08}. 

Regarding the ANN, our implementation relies on the Matlab Neural Network toolbox. The ANN is a feedforward network consisting of a single hidden layer consisting of neurons (fixed to 16 in this work) utilising a $\tanh$ transfer function.  The output layer is a simple neuron with a logarithmic sigmoid function suited for a binary decision. The training is conducted using a standard error back-propagation algorithm \citep{PRML}. Although not strictly true the ANN output is treated as a posterior probability in this study.


\subsection{Experimental Protocol}
\label{ssec:ExpProtocol}

\subsubsection{Metric}

To assess the performance of the algorithms, we use the F1 score as a frame-level metric. This measure combines true positives (Tp), false positives (Fp) and false negatives (Fn) into one single metric. This metric is particularly useful when analysing skewed datasets where the feature being considered has a rather sparse occurrence (e.g., for laughter detection, \citealp{scherer09}), and is therefore well suited for assessing the performance of creak detection techniques. The metric is bound between 0 and 1, with 1 indicating perfect detection:

\begin{equation}
 \label{eq:F1}
F1 = \frac{2\cdot\mbox{Tp}}{2\cdot\mbox{Tp} + \mbox{Fp} + \mbox{Fn}} \in [0, 1]
\end{equation}

\subsubsection{Threshold setting}
Once the ANN and BDT based classifiers have been trained it is then possible to set the decision threshold, $\alpha$, in order to optimise performance on the training set (the splitting of training and test sets is outlined in Section \ref{sec:xval} below). Note that the classifiers output the posterior probability, $P_1$, of a given sample corresponding to class 1 (i.e. creaky). The standard binary decision is class 1 if $P_1 > \alpha$ (otherwise class 0) and typically $\alpha$ is set to 0.5. However, for skewed datasets which contain a given class to be detected which displays sparse occurrence, this setting for $\alpha$ may not be optimal. In the present study, we systematically vary $\alpha$ in the range [0, 1], and set it to the value which maximises the F1 score on the training set. This decision threshold is then applied during subsequent testing. Note that we observed this threshold setting to have a very low inter-database sensitivity, as all speakers had their best F1 score for $\alpha$ in the 
vicinity of 0.3. 


\subsubsection{Cross-validation}
\label{sec:xval}
In order to evaluate the detection performance of the various methods, analysis was carried out on the speech databases described in Section \ref{sec:databases}. A leave one speaker out design was used whereby the speech data of a given speaker was held out for testing and the remainder of the speech data was used for training the classifier. This procedure was repeated for each speaker.

\subsection{Results}
\label{sec:DetectionResults}

\begin{figure}[ht!]
  \begin{center}
   \includegraphics[width=14cm]{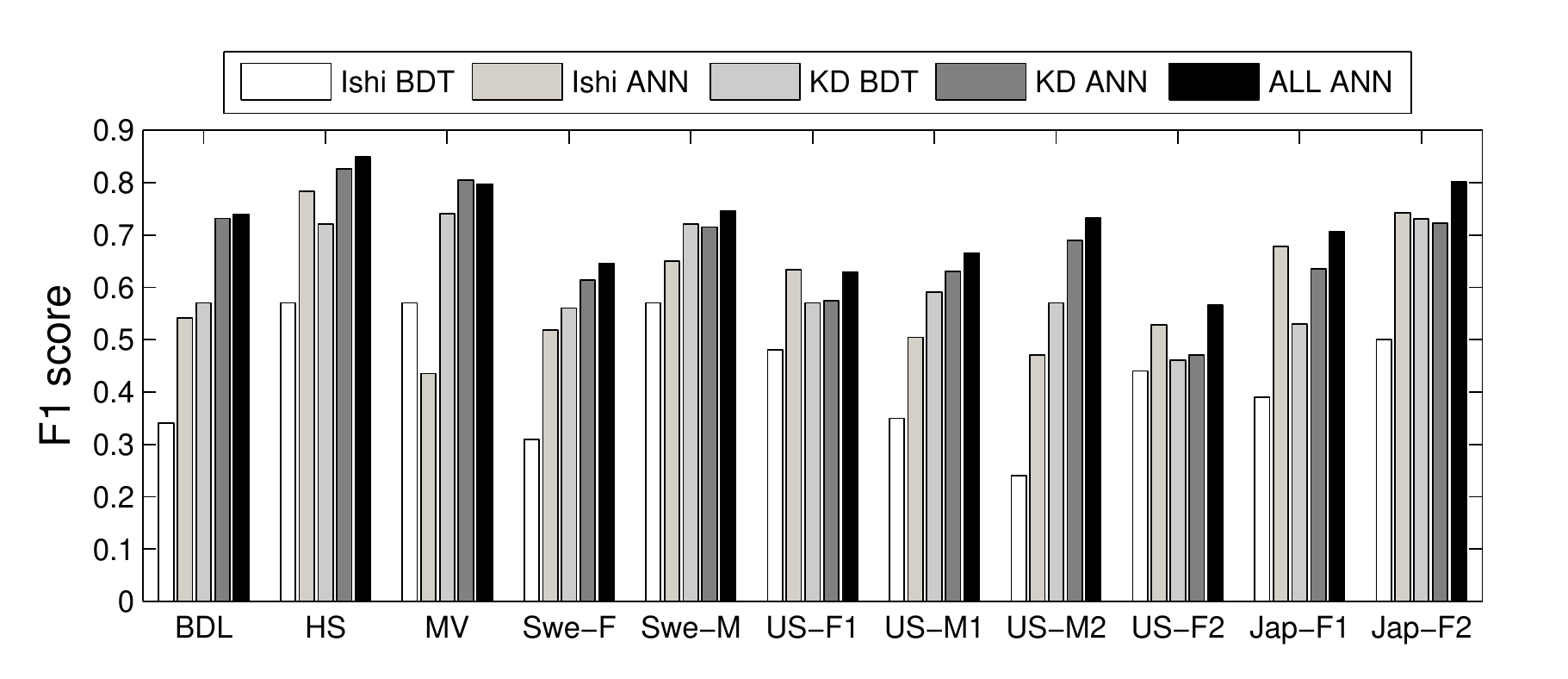}
  \end{center}
\vspace{-.6cm}
\caption{Automatic detection results, summarised as F1 score, across the various speakers.} 
\label{fig:DetectionResults}
\vspace{0cm}
\end{figure}

Our detection results are summarised in Figure \ref{fig:DetectionResults} for all methods and across all datasets. Note that the \emph{KD-BDT} system is the one we proposed in a previous paper \citep{kane13creakCSL}. Firstly, the influence of the classifier (BDT vs. ANN) is analysed. Out of the 22 comparisons (11 speakers and two feature sets), it turns out that BDT is outperformed by ANN in all but one occasion. This finding was supported by evidence from a two-way ANOVA (with F1 score treated as dependent variable, and classifier type and feature group as the two independent variables) which revealed a highly significant effect of classifier type [$F_{(1, 50)} = 23.0, p < 0.001$]. As a consequence, when considering the full feature set, only the ANN was used as a classifier. The superiority of the ANN classifier over BDT is likely due to its ability to utilise the interaction of the various features in discriminating creaky voice from other speech regions. 

One exception where the ANN classifier is actually outperformed is in the case of the Ishi's features for speaker MV. MV actually has a particularly low-pitched voice, with an average f0 in the vicinity of 80 Hz. The ANN trained classifier with Ishi's features produces a large number of false positives with many non-creaky low-pitched regions being mistakenly determined as creaky. We speculate that the small number of Ishi's features are actually over-fit to speakers who are not particularly low-pitched using the ANNs (as this type of speech is not very well represented in the training data) whereas with this setup the decision trees do not over-fit to the same extent.

Regarding the relevance of the features, results in Figure \ref{fig:DetectionResults} corroborate our conclusions from Section \ref{sec:relevance}: the KD features are more informative for the discrimination of creaky phonation. In general, they achieved much higher F1 scores than when using Ishi's features. This finding is supported by evidence from the same two-way ANOVA testing mentioned above, where a significant effect of feature group was observed [$F_{(2, 50)} = 11.6, p < 0.001$]. Further, a subsequent pairwise comparison using Tukey's Honestly Significant Difference (HSD) test indicated a significant difference (p $<$ 0.001) between the KD and Ishi feature groups.

The combination of both feature sets (\emph{All-ANN}) leads to the best overall detection results, with the Tukey HSD test revealing a significant improvement compared to Ishi-ANN (p $<$ 0.05). Note, however, that although the mean F1 score is higher for All-ANN compared with KD-ANN the difference does not achieve significance (p = 0.92).
Notice also that compared to our previous method proposed in \cite{kane13creakCSL} (\emph{KD BDT}), a substantial increase of performance has been achieved. 

One can observe a certain degree of between speaker varibility in detection performance. In particular, for speaker US-F2, the best classifier (i.e. All-ANN) produces the lowest F1 score (0.57) for any speaker. For this speaker, it was noticed that the audio contained sounds resulting from collisions of the microphone with the speaker's mouth, and this led to a relatively high volume of false detections.

For the interested researcher, we have made the resulting \emph{All ANN} algorithm freely available on the web \footnote{GLOAT toolkit:  \url{http://tcts.fpms.ac.be/~drugman/Toolbox/}, the  Voice Analysis Toolkit: \url{http://www.tcd.ie/slscs/postgraduate/phd-masters-research/student-pages/johnkane.php}, and the COVAREP project: \url{https://github.com/covarep/covarep}}.

\begin{figure}[ht!]
  \begin{center}
   \includegraphics[width=15cm]{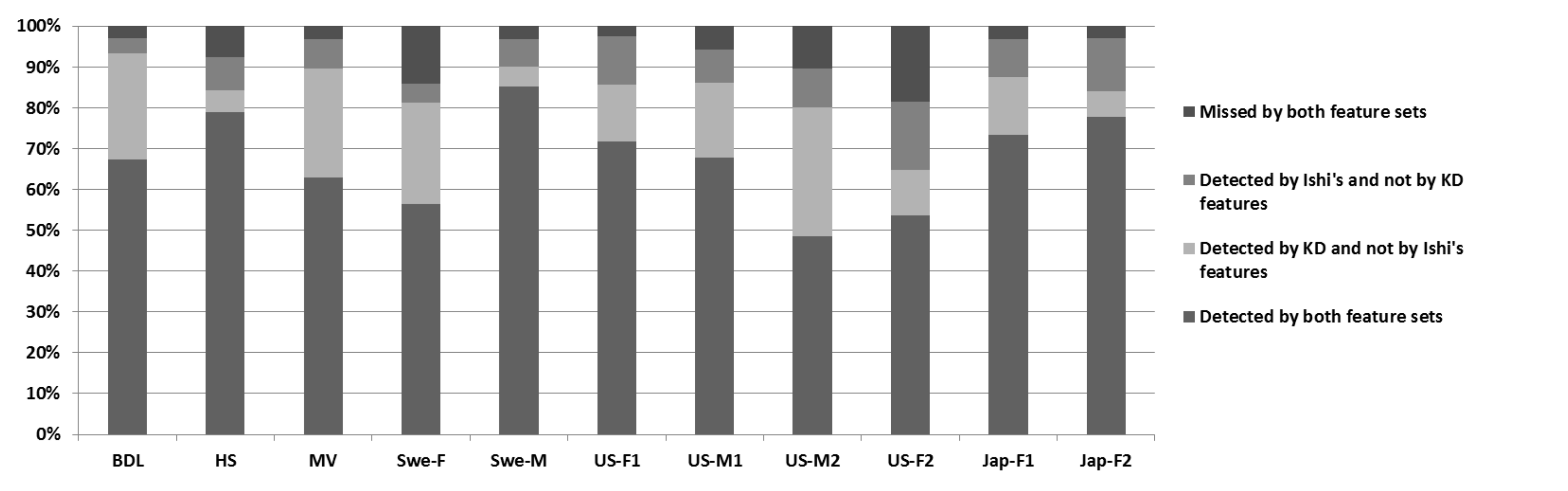}
  \end{center}
\vspace{-.6cm}
\caption{For each speaker, repartition of the events: \emph{i}) missed when using either KD or Ishi's features, \emph{ii}) detected when using Ishi's features but missed using KD features, \emph{iii}) detected when using KD features but missed using Ishi's features, \emph{iv}) detected when using either KD or Ishi's features. } 
\label{fig:DetectionAnalysis}
\vspace{0cm}
\end{figure}

In order to confirm the possible presence of various patterns in creaky voice, and based on our findings from Section \ref{sec:relevance}, we inspected whether creaky events have been detected or not using Ishi's features (\emph{Ishi-ANN} system) or using our KD features (\emph{KD-ANN} system). Each event can be assigned to the following categories: i) it is missed by both systems, ii) it is detected using Ishi's features but not using KD ones, iii) it is detected using KD features but not using Ishi's ones, iv) it is detected by both systems. Results across all datasets are exhibited in Figure \ref{fig:DetectionAnalysis}. It can be first observed that missing rates are generally low (below 10\% for 9 out of the 11 speakers). Note that the false alarm rates are not explicitly reported in this paper (though of course they significantly affect the F1 metric), but we noticed that a good trade-off between misses and false alarms was obtained for the proposed method. Therefore, the false alarm rates are 
comparable to 
the miss rates achieved in Figure \ref{fig:DetectionAnalysis}. 

Secondly, it is interesting to see how Figure \ref{fig:DetectionAnalysis} corroborates our results from Figure \ref{fig:FinalMIProfile}. It can be indeed seen that speakers for whom MI values were high for KD features and lower for Ishi's features have an important proportion of events detected using KD features but not using Ishi's ones (while the contrary is not true), and vice versa. This provides a further experimental support which consolidates the existence of at least two patterns that speakers might use to produce creaky voice. An investigation into these patterns will be given in Section \ref{sec:patterns} through an analysis of the excitation signal.

\section{Creaky Patterns Analysis}
\label{sec:patterns}
Our findings from Sections \ref{sec:relevance} and \ref{sec:detection} tend to indicate that the production of creaky voice does not involve an unique strategy and that speakers use at least two creaky excitation patterns (corroborating previous findings in the literature; \citealp{redi01,ishi10,ishi08,ishi07}). Based on the events exclusively detected using either Ishi's or KD features, we visually inspected a large number of creaky excitation segments from our 11 speakers. From this analysis, it turned out that creaky excitation signals could be categorised into three patterns: highly irregular temporal characteristics, fairly regular temporal characteristics with strong secondary excitation peaks and fairly regular temporal characteristics without strong secondary excitations. Note that the first pattern closely corresponds to the `multi-pulsed' \citep{ishi08} or first category \citep{redi01} patterns previously reported. The second and third patterns are essentially a subdivision of the `single-pulsed' 
or second 
category previously reported patterns.

A more complete description of these three patterns follows in Section \ref{sec:Qualitative}. In this description we focus on the LP-residual waveform as this signal, in the present context, can reveal some important characteristics not obvious from the speech signal and also as our research is concerned with the modelling of the residual excitation patterns of speech for statistical parametric speech synthesis \citep{drugman12DSM,drugkane12_2}. A quantitative study of the identified creaky patterns is eventually provided in Section \ref{sec:Quantitative}.

\subsection{Qualitative Analysis of the Identified Creaky Patterns}
\label{sec:Qualitative}

Three patterns have been identified across our 11 speakers. There are now thoroughly described.

\subsubsection{Pattern A: Highly irregular temporal characteristics}
\label{sec:pat1}

\begin{figure}[ht!]
  \begin{center}
   \includegraphics[width=12cm]{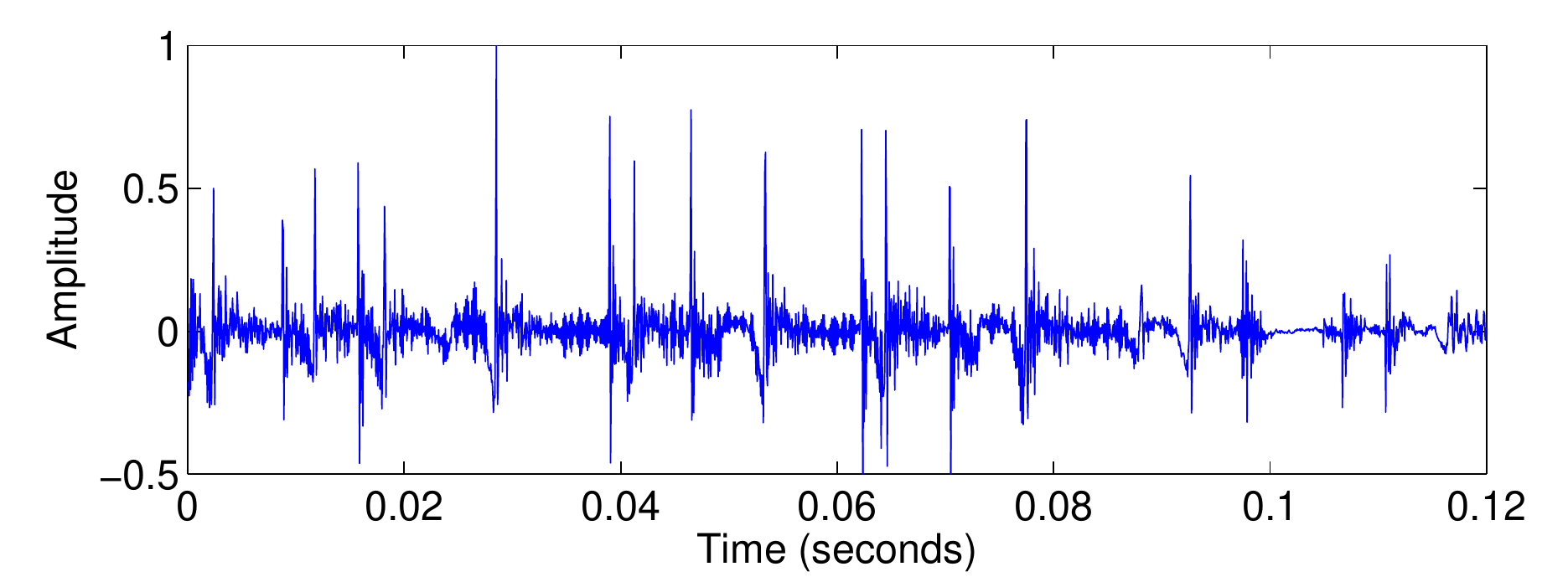}
  \end{center}
\vspace{0cm}
\caption{LP-residual of a segment of a speech signal containing creaky voice with irregular temporal characteristics.} 
\label{fig:patt1}
\vspace{0cm}
\end{figure}

An illustration of Pattern A is displayed in Figure \ref{fig:patt1}. During such a creaky production, the LP residual signal exhibits clear discontinuities with a highly irregular temporal structure. These peaks appear sporadically, and the inter-peak duration does not seem to follow a clear deterministic rule as it is the case for regular patterns. Often occuring within this pattern are spells of diplophonia, with amplitude and duration variation in successive pulses. Note, however, that a stable diplophonic pattern (such as that described in \citealp{klatt90}) was not observed to be sustained throughout a creaky voice segment. In the case of such a creaky production, the individual KD features turn out to be somewhat erratic: \textit{i)} since the pattern is very irregular, the $F0$ and subsequent \emph{H2-H1} parameter extraction becomes less straightforward; \textit{ii)} excitation peaks are not prominent at the scale used for the calculation of \emph{Peak-Prom}, and further because often a 
large proportion of the glottal pulses displays a duration corresponding to an $F0$ higher than 70 Hz. Contrastingly, the combination of the intra-frame periodicity (IFP) and inter-pulse similarity (IPS) parameters proposed by \cite{ishi08} are effective for the detection of such a 
phonation as IFP values become very close to 0 in such regions, 
compared to the higher values (closer to 1) in more regular phonatory regions, while IPS values stay high in both regions.

\subsubsection{Pattern B: Fairly regular temporal characteristics with strong secondary excitation peaks}
\label{sec:pat2}

\begin{figure}[ht!]
  \begin{center}
   \includegraphics[width=12cm]{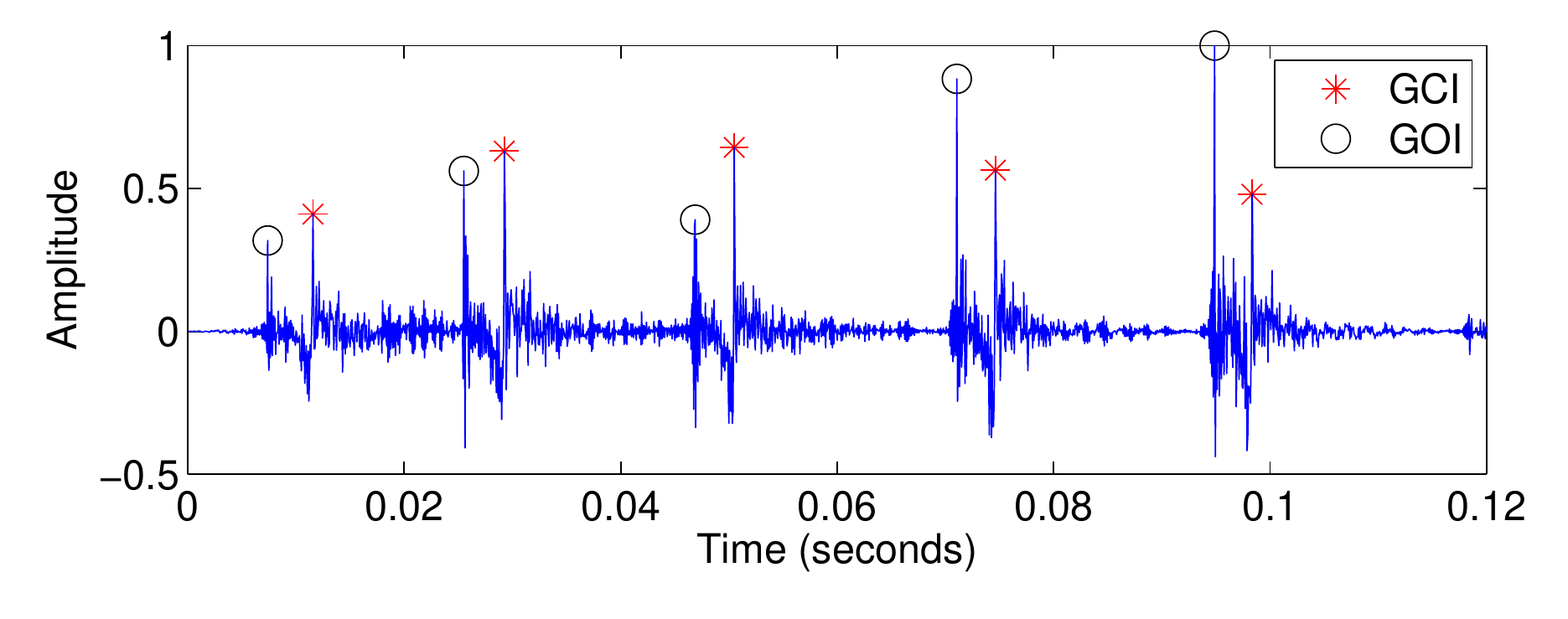}
  \end{center}
\vspace{0cm}
\caption{LP-residual of a segment of a speech signal containing creaky voice with very long glottal pulses, reasonably regular temporal characteristics and strong secondary peaks. Peaks corresponding to glottal closure instants are marked with a red star while glottal opening instant peaks are marked with a black circle.} 
\label{fig:patt2}
\vspace{0cm}
\end{figure}

An example of Pattern B is displayed in Figure \ref{fig:patt2}. It can be observed that the LP residual signal exhibits fairly regular temporal characteristics, with a stable pattern comprising two clear discontinuities. Note that we are not claiming that these patterns display strictly the same extent of regular temporal characteristics as modal voice, but simply that they are considerably more regular than those characteristics in Pattern A. In this figure, these two peaks have been annotated for each glottal cycle: the peak indicated by a red star corresponds to the Glottal Closure Instant (GCI), while the peak indicated by a circle is called the secondary excitation peak. Note that the SE-VQ algorithm \citep{kaneGCI} which is an extension of the SEDREAMS method \citep{drugman12GCI} is used for detecting GCIs and that secondary peaks are simply measured as the strongest peak between adjacent GCIs. These secondary peaks (black circles) sometimes occur due to secondary laryngeal excitations, but very often 
they stem from sharp discontinuities at glottal opening, following a long glottal closed period.  This is in fact the case in Figure  \ref{fig:patt2} and this observation was supported by evidence from both the corresponding EGG signal as well as the glottal source derivative estimated by glottal inverse filtering. This appears to be consistent with the ``double-beated'' pattern observed in \cite{ishi07}.

\begin{figure}[ht!]
  \begin{center}
   \includegraphics[width=14cm]{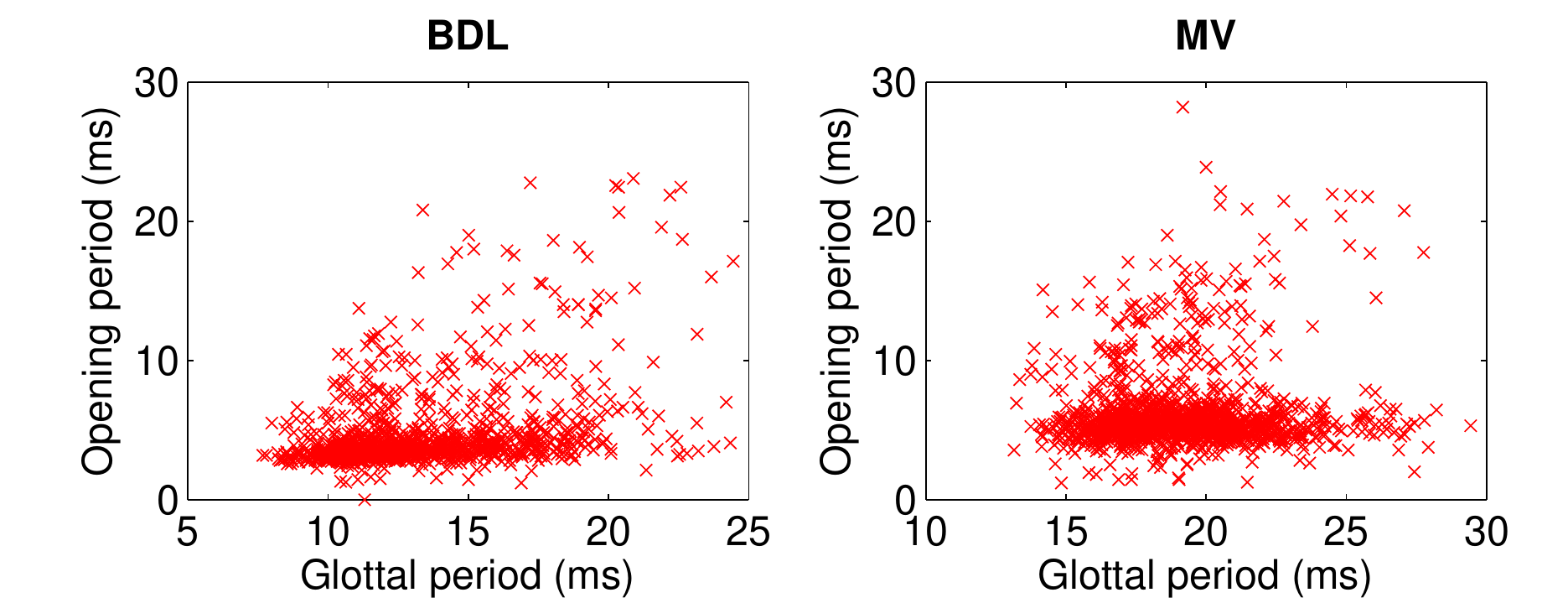}
  \end{center}
\vspace{0cm}
\caption{Distributions, for two male speakers who mostly used pattern B, of the glottal opening period as a function of the produced pitch period.} 
\label{fig:patt2_scatter}
\vspace{0cm}
\end{figure}

The two speakers whose creaky voice regions were detected mainly due to the KD features and not Ishi's features (refer to Figure \ref{fig:FinalMIProfile}), BDL and MV, were found to mostly display this creaky voice pattern. For these speakers, considering the duration of the glottal periods against the duration of the glottal opening phase (i.e. duration from a secondary peak to the consecutive GCI) shown in Figure \ref{fig:patt2_scatter}, one can observe the striking trend of a relatively stable glottal opening period despite a widely varying glottal period. This indicates that the glottal closed phase is the main determinant of the glottal period (and hence $F0$), and confirms our observation from \citep{drugkane12_2}. Note that the detection of GCIs and GOIs for this type of phonation is often extremely difficult \citep{kaneGCI} and consequently spurious values will inevitably have occurred in the analysis which would have influenced these plots (i.e. Figure \ref{fig:patt2_scatter}).

Both the very prominent secondary excitation peaks and this trend observed in Figure \ref{fig:patt2_scatter} likely have a significant impact on the perceptual quality of these creaky voice patterns. Indeed we have found that for these two speakers, considering these features in the design 
of an excitation model brought about a clear improvement 
in the rendering of the voice quality in the synthesis experiments reported in \cite{drugkane12_2}.

\subsubsection{Pattern C: Fairly regular temporal characteristics without strong secondary excitation peaks}
\label{sec:pat3}
\begin{figure}[ht!]
  \begin{center}
   \includegraphics[width=12cm]{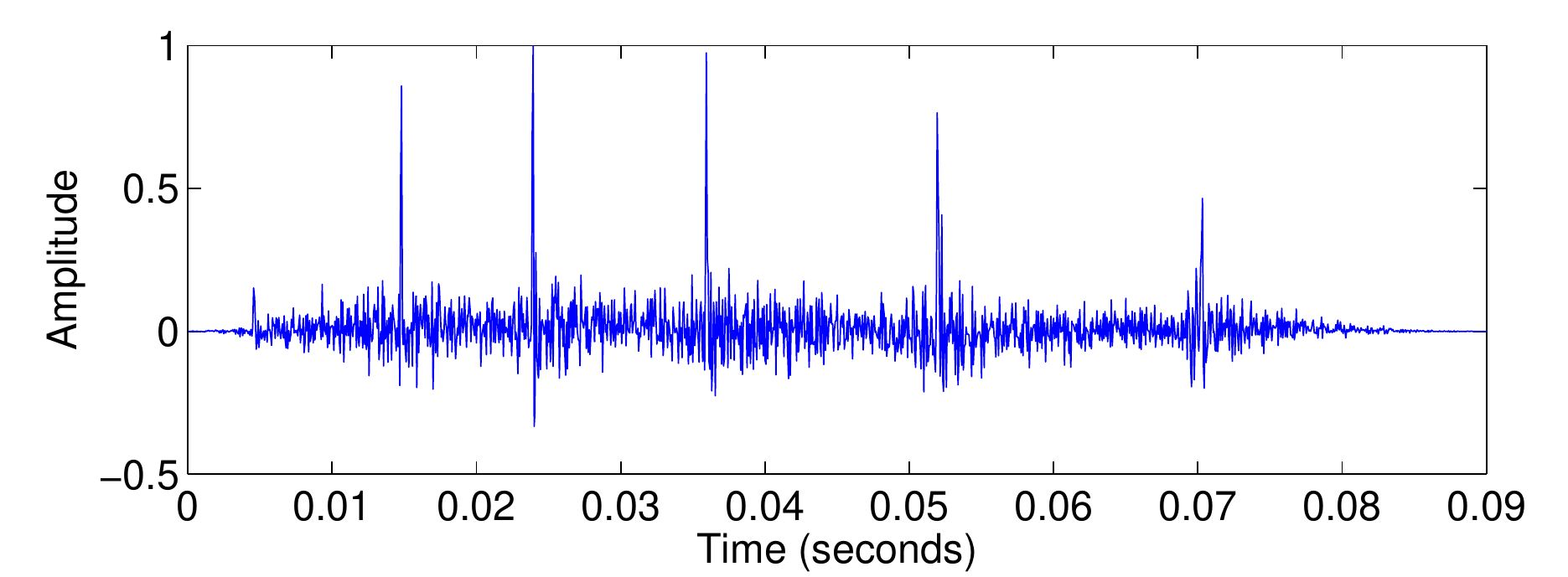}
  \end{center}
\vspace{0cm}
\caption{LP-residual of a segment of a speech signal containing creaky voice with very long glottal pulses, reasonably regular temporal characteristics and no strong secondary peaks.} 
\label{fig:patt3}
\vspace{0cm}
\end{figure}

The third pattern of creaky voice observed in the present study (an example of which is shown in Figure \ref{fig:patt3}), similarly to Pattern B, likely corresponds to the `single-pulsed' \citep{ishi08} or the second category \citep{redi01} pattern previously reported in the literature \citep{redi01,ishi08}. As with Pattern B, the temporal characteristics observed here are fairly regular, particularly in comparison to Pattern A. The perceptual effect of creaky voice is mainly brought about here by the low $F0$. However, unlike Pattern B, Pattern C does not display strong secondary excitation peaks. The consequence of this for speech synthesis is that the modelling of this excitation pattern can be carried out in a similar fashion to modal phonation, and its proper rendering is conditioned mainly on the generation of a suitable $F0$ contour.

\subsection{Quantitative Analysis of the Identified Creaky Patterns}
\label{sec:Quantitative}

In the previous sections, we have identified three main patterns of creaky voice. These patterns have been described and qualitatively analyzed. We now perform a quantitative study by investigating how frequent these patterns are observed, what their speaker dependency is and how effective the systems of Section \ref{sec:detection} are at detecting these patterns. For this, the same portion of the corpus introduced in Section \ref{sec:databases} was annotated by the first two authors. This dataset consists of 5 minutes of speech for each of the 11 speakers. The speech and residual waveforms of each creaky segment (which was manually labelled as described in Section \ref{sec:annot}) were presented to the annotator. Based on this visualization, the annotator had to assign the creaky segment one (and only one) of the three identified creaky patterns, strictly following the criteria described in Section \ref{sec:Qualitative}.

The inter-annotator agreement rate was estimated using the Krippendorff's alpha \citep{Krippendorff}. Our results report a Krippendorff's alpha value of 0.718 which indicates a reasonably high level of inter-rater agreement. The confusion matrix between the 3 patterns is shown in Table \ref{tab:ConfMatrix}. Thre greater risk of confusion concerns the pairs (A,B) and (A,C). This is because some cases were particularly ambiguous as their degree of regularity was neither totally regular nor totally irregular, which inherently induces to some subjectivity issues and might explain the discrepancies across annotators.

\begin{table}[!ht]
\centering
\begin{tabular}{ c|ccc}
\toprule
 & \textbf{Pattern A} & \textbf{Pattern B} & \textbf{Pattern C}\\
\midrule   
\textbf{Pattern A} & 97.66 & 1.87 & 0.47\\
\textbf{Pattern B} & 33.22 & 65.37 & 1.41\\
\textbf{Pattern C} & 19.42 & 2.16 & 78.42\\
\bottomrule
  
\end{tabular}
\caption{Confusion matrix (in \%) for the three identified creaky patterns across the two annotators.}
\label{tab:ConfMatrix}
\end{table}

\begin{figure}[ht!]
  \begin{center}
   \includegraphics[width=13cm]{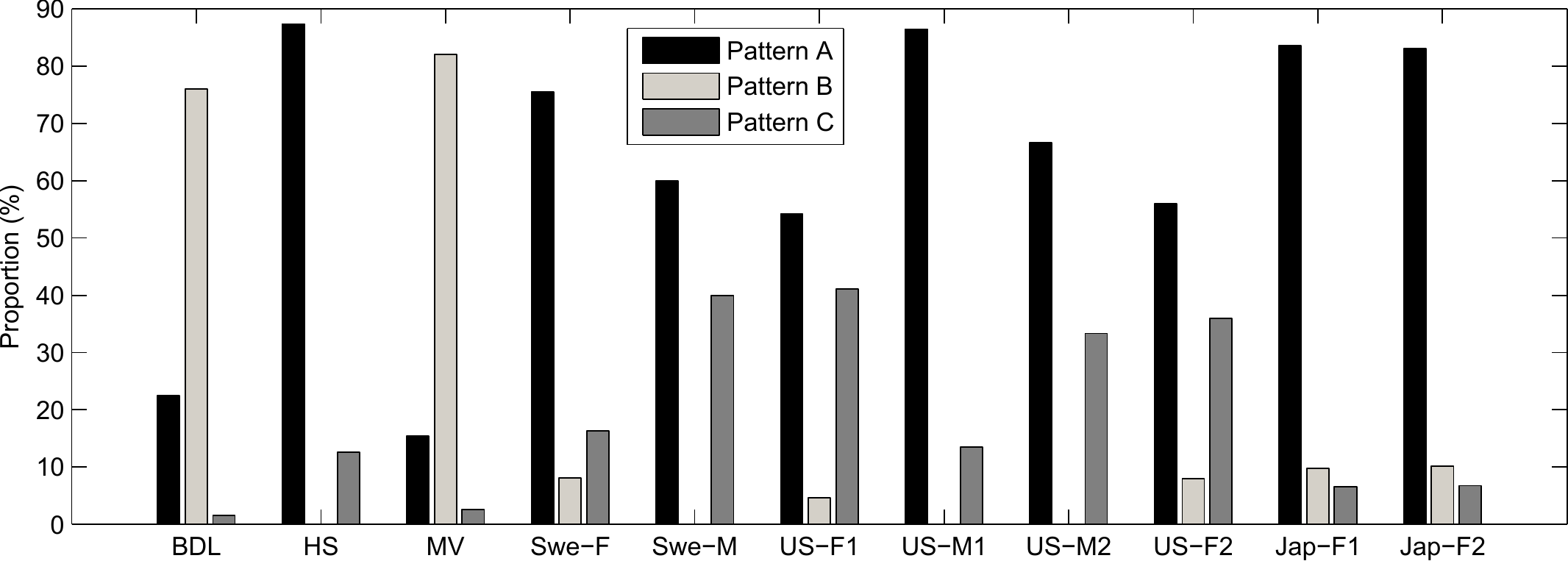}
  \end{center}
\vspace{-.6cm}
\caption{Repartition (in \%) of the creaky pattern usage for our 11 speakers.} 
\label{fig:PatternRepartition}
\vspace{0cm}
\end{figure}

Based on these new annotations, we have first investigated how frequent do speakers use the different patterns. For this, we only consider the manual labels for which both annotators agreed. The repartition of the creaky usage per speaker is displayed in Figure \ref{fig:PatternRepartition}. First of all, a considerable speaker dependency can be observed. One first conclusion is that pattern A is the most frequently used pattern, except for male speakers BDL and MV whose dominant pattern is clearly B. This is an important observation as the creaky vocoder system we have developed in \citep{drugkane12_2}, and which was further integrated into HMM-based speech synthesis in \citep{raitioCreakSynth}, is based on the modeling of pattern B. This is because in these latter studies, only voices from the TTS set have been considered (BDL, HS and MV) and pattern B appeared to be the most frequent characteristic of creaky voice. \textit{A posteriori}, this seems to be biased as pattern A turns out to be ubiquitously used in 9 
out of our 11 speakers. 

A second observation is that although a great inter-speaker variability is noticed in Figure \ref{fig:PatternRepartition}, speakers could be categorized into similar groups:
\begin{itemize}
\item BDL and MV who predominantly use pattern B,
\item HS, Swe-F, US-M1, Jap-F1 and Jap-F2 who use pattern A with more than 75\%, the rest being shared between B and C (B being sometimes even not used at all),
\item Swe-M, US-F1, US-M2 and US-F2 who use pattern C between 33 and 42\%, and pattern A almost exclusively in other cases.
\end{itemize}

As a consequence, it is essential that a parametric speech synthesizer of creaky voice incorporates a proper modeling of the irregularities of pattern A. Note that the modeling of pattern B was already proposed in \citep{drugkane12_2} and that the excitation in pattern C is very similar to the excitation signals in modal voice but is merely characterized by much lower $F_0$ values. This latter could therefore be synthesized using any existing excitation modeling. 

Finally, we address how the performance of the creaky voice detection systems developed in Section \ref{sec:detection} is affected across the three identified creaky patterns. For this, we run classification experiments over the patterns where both annotators agreed. It is worth emphasizing at this point that our goal is not to build up classification systems able to discriminate between the different creaky patterns, but rather to investigate how the systems of Section \ref{sec:detection} perform for each creaky pattern. As a performance measure, we here use miss rate (also called \emph{false negative error} rate).  Indeed, metrics based on false alarm measurements (and also for F1 score) would be meaningless here, since these false alarms could not be assigned to a specific pattern. 

\begin{figure}[ht!]
  \begin{center}
   \includegraphics[width=12cm]{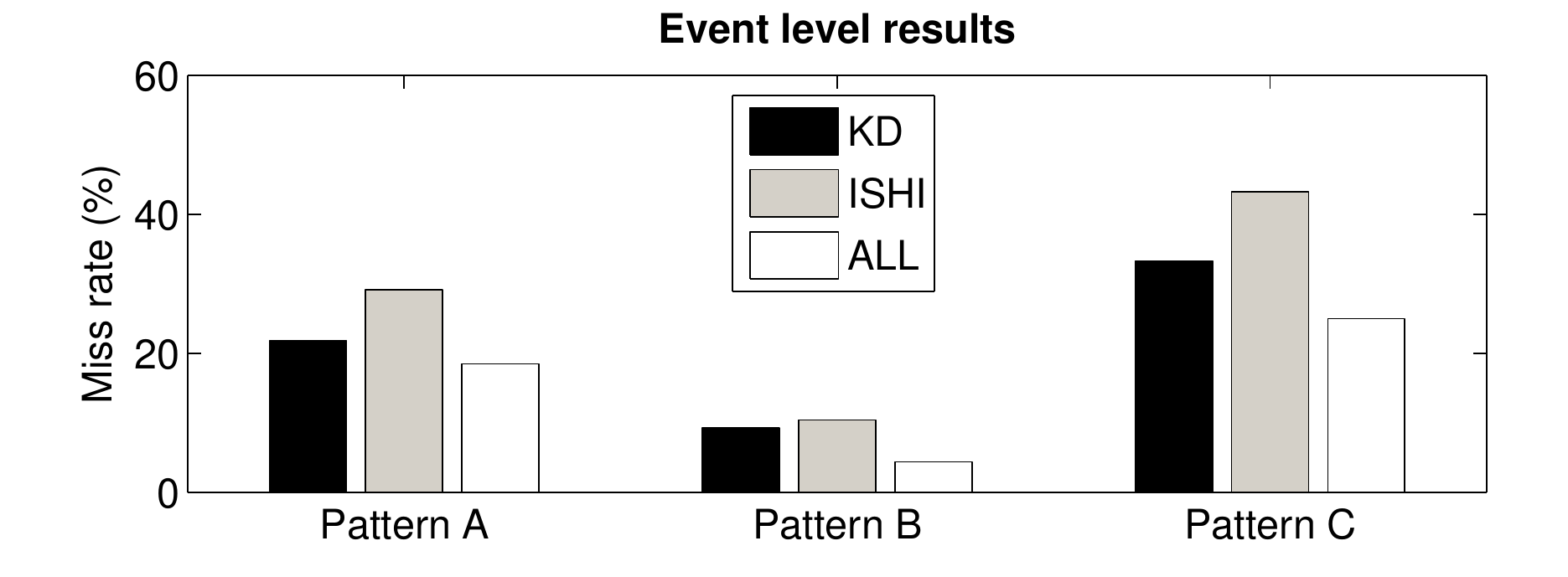}
  \end{center}
\vspace{-.6cm}
\caption{Classification miss rate results across the 3 identified creaky patterns.} 
\label{fig:ClassificationPerPattern}
\vspace{0cm}
\end{figure}

Results are shown in Figure \ref{fig:ClassificationPerPattern} for the 3 ANN-based systems using either \textit{Ishi}, \textit{KD} or \textit{All} feature sets. Across all three patterns there is a consistent trend, with All features achieving the lowest classification miss rate, KD features with the second lowest and Ishi features with the highest. There are also considerable inter-pattern differences. 
Interestingly for pattern A, which involves a highly irregular temporal excitation pattern, KD features result in a lower miss rate compared to the Ishi features classifier. Despite the individual KD features displaying relatively erratic contours in these regions, their combination within the ANN classifier achieves effective detection of this pattern. Pattern B, which contains excitation patterns with strong secondary peaks, is rarely missed by the three classifiers. The highest overall miss rates are observed for Pattern C.

\section{Conclusion}
\label{sec:concl}
This paper addressed the automatic analysis of the excitation patterns used during the production of creaky voice. The goal was threefold: \emph{i)} to assess the relevance of acoustic features specifically designed to characterise creaky voice; \emph{ii)} to integrate these combined features within an efficient creaky voice detection system; \emph{iii)} based on these results, to categorize the observed temporal creaky patterns, and to analyse them both qualitatively and quantitatively.

This study was led on a large amount of manually-annotated data with a variety of languages, speakers and recording conditions (read vs conversational speech). Our analysis was based on the use of acoustic features which were previously specifically designed for the characterisation of creaky voice. Our approach consisted of three consecutive stages. In the first one, the relevance of each feature individually as well as their complementarity/redundancy were assessed relying on mutual information-based measures. Two groups of features were interestingly noticed to be complementary: features proposed by Ishi and in one of our previous studies. This is initial evidence supporting the notion of the presence of several creaky patterns as it shows that these features describe different characteristics of creaky voice. In the second step, these features were the input of two classifiers (a binary 
decision tree and an artificial neural network) with the goal of automatic creaky voice detection. 

Two conclusions were drawn from our classification experiments. First the use of the combined feature set with neural networks led to an appreciable improvement over the state of the art across our 11 speakers. Secondly, an analysis of the detected creaky events strengthened our findings about the evident existence of several creaky patterns. Finally, in the last stage of our approach, we inspected a large number of creaky excitation signals detected exclusively using one of the two sets of features, as supported by the two first experiments. 

Our analysis revealed 3 distinct creaky excitation patterns. In pattern A, the LP-residual signal exhibits important discontinuities occurring sporadically and whose temporal structure does not seem to follow any obvious deterministic rule. Contrastingly, the temporal characteristics in patterns B and C is much more regular and often highly periodic.
In both patterns, the discontinuity at the glottal closure instants (GCIs) is well marked, and the glottal period (which separates two consecutive GCIs) is much larger than in modal speech. In the case of pattern B, the LP-residual displays one extra peak per glottal cycle in addition to the discontinuity at GCIs. These secondary excitation peaks are often likely due to the sudden opening of the glottis. Interestingly, we observed for two speakers who mostly used pattern B, that the opening period remained almost constant independently of the produced pitch. Finally, these secondary peaks were not observed in the case of pattern C whose characteristics are similar to the excitation used in modal voiced phonation, but with $F0$ values falling below 50 Hz.

A quantitative analysis of the identified creaky patterns was carried out. Three main conclusions could be drawn from our experiments: \emph{i)} a Krippendorff's alpha value of 0.718 was obtained when annotating the patterns, which indicates a reasonably high level of inter-rater agreement; \emph{ii)} a considerable inter-speaker variability was observed in the way the creaky patterns are used; \emph{iii)} the ANN-based classifier using both Ishi's and our proposed features misses creaky events in the following proportion: 18.5\% for pattern A, 4.4\% for pattern B and 25\% for pattern C.

For future work, it would be worth further investigating the effect of different classifier settings. Such analysis could involve optimising classifier parameters (e.g., number of neurons in the hidden layer), assessing different classifier types and also determining the value of fusing multiple classifiers.  We intend to utilise the new detection algorithm as part of our on-going developments incorporating creaky voice in speech synthesis and voice quality transformation of synthetic vocies. Indeed, the vocoder we have developed in \citep{drugkane12_2} models only pattern B. However, our findings in the present study, in particular that pattern A is dominantly used by 9 out of the 11 speakers, suggest that further effort needs to be invested into modelling the highly irregular excitation characteristics of Pattern A. It is therefore essential to develop an excitation model able to model the temporal irregularities as found in pattern A. It is also envisaged that the algorithm will  be used to study the use of creaky voice in conversational and expressive speech. Furthermore, as creaky voice is thought to be associated with 
certain affective states, its automatic detection may bring significant benefit to applications like emotion recognition. Another direction of future work could involve carrying out analysis-synthesis experiments to help better determine the acoustic characteristics required for the 
perception of creaky voice.

\section{Acknowledgements}
\label{sec:acknow}
The first author is supported by FNRS. The second and third authors are supported by the Science Foundation Ireland Grant 09 / IN.1 / I 2631 (FASTNET) and the Irish Department of Arts, Heritage and the Gaeltacht (ABAIR project). The authors would like to thank Jens Edlund for providing us with the Spontal corpus, Ikuko Patricia Yuasa for providing us with the US dataset and Thomas Magnuson for sharing the Japanese data with us. 

\bibliographystyle{elsarticle-harv}

\end{document}